\documentclass[final]{IEEEtran}

\usepackage{extarrows}
\usepackage{graphicx}
\usepackage{subfigure}
\usepackage{multirow}
\usepackage{array}
\newcolumntype{P}[1]{>{\centering\arraybackslash}p{#1}}
\newcolumntype{M}[1]{>{\centering\arraybackslash}m{#1}}

\usepackage{amsthm,amssymb,amsmath,graphicx,multirow,color,amsfonts}%
\usepackage[update,prepend]{epstopdf}
\usepackage{multirow}
\usepackage[latin1]{inputenc}
\usepackage{tikz}
\usepackage{bbm} 
\usepackage{pdfpages}
\usepackage{multirow}
\usepackage{subfig}
\usepackage{comment}

\usepackage{graphicx}
\usepackage{multicol}
\usepackage{cite}

\usepackage[justification=centering]{caption}
\usepackage{textcomp}
\usepackage{psfrag}
\usepackage{arydshln}
\usepackage{url}
\usepackage{soul}
\usepackage{graphicx,color}
\usepackage[nolist]{acronym}
\usepackage{algorithm,algorithmic} 

\usepackage{mathtools,lipsum}
\usepackage{cuted}
\usepackage{amsmath}
\newtheorem{proposition}{Proposition}
\usepackage{mathrsfs}

\usepackage[capitalise]{cleveref}
\Crefname{equation}{Eq.\!}{Eqs.\!}
\Crefname{figure}{Fig.\!}{Figs.\!}
\Crefname{tabular}{Tab.\!}{Tabs.\!}
\Crefname{section}{Section\!}{Sections.\!}

\def\nb0{{\mathbf{0}}}
\def\nb1{{\mathbf{1}}}

\newtheorem{lemma}{Lemma}

\newtheorem{definition}{Definition}

\newtheorem{theorem}{Theorem}

\def\argmin{\operatorname{arg~min}}

\begin{document}
\graphicspath{{./Figures/}}
\begin{acronym}

\acro{5G-NR}{5G New Radio}
\acro{3GPP}{3rd Generation Partnership Project}
\acro{ABS}{aerial base station}
\acro{AC}{address coding}
\acro{ACF}{autocorrelation function}
\acro{ACR}{autocorrelation receiver}
\acro{ADC}{analog-to-digital converter}
\acrodef{aic}[AIC]{Analog-to-Information Converter}     
\acro{AIC}[AIC]{Akaike information criterion}
\acro{aric}[ARIC]{asymmetric restricted isometry constant}
\acro{arip}[ARIP]{asymmetric restricted isometry property}

\acro{ARQ}{Automatic Repeat Request}
\acro{AUB}{asymptotic union bound}
\acrodef{awgn}[AWGN]{Additive White Gaussian Noise}     
\acro{AWGN}{additive white Gaussian noise}

\acro{APSK}[PSK]{asymmetric PSK} 

\acro{waric}[AWRICs]{asymmetric weak restricted isometry constants}
\acro{warip}[AWRIP]{asymmetric weak restricted isometry property}
\acro{BCH}{Bose, Chaudhuri, and Hocquenghem}        
\acro{BCHC}[BCHSC]{BCH based source coding}
\acro{BEP}{bit error probability}
\acro{BFC}{block fading channel}
\acro{BG}[BG]{Bernoulli-Gaussian}
\acro{BGG}{Bernoulli-Generalized Gaussian}
\acro{BPAM}{binary pulse amplitude modulation}
\acro{BPDN}{Basis Pursuit Denoising}
\acro{BPPM}{binary pulse position modulation}
\acro{BPSK}{Binary Phase Shift Keying}
\acro{BPZF}{bandpass zonal filter}
\acro{BSC}{binary symmetric channels}              
\acro{BU}[BU]{Bernoulli-uniform}
\acro{BER}{bit error rate}
\acro{BS}{base station}
\acro{BW}{BandWidth}
\acro{BLLL}{ binary log-linear learning }

\acro{CP}{Cyclic Prefix}
\acrodef{cdf}[CDF]{cumulative distribution function}   
\acro{CDF}{Cumulative Distribution Function}
\acrodef{c.d.f.}[CDF]{cumulative distribution function}
\acro{CCDF}{complementary cumulative distribution function}
\acrodef{ccdf}[CCDF]{complementary CDF}               
\acrodef{c.c.d.f.}[CCDF]{complementary cumulative distribution function}
\acro{CD}{cooperative diversity}

\acro{CDMA}{Code Division Multiple Access}
\acro{ch.f.}{characteristic function}
\acro{CIR}{channel impulse response}
\acro{cosamp}[CoSaMP]{compressive sampling matching pursuit}
\acro{CR}{cognitive radio}
\acro{cs}[CS]{compressed sensing}                   
\acrodef{cscapital}[CS]{Compressed sensing} 
\acrodef{CS}[CS]{compressed sensing}
\acro{CSI}{channel state information}
\acro{CCSDS}{consultative committee for space data systems}
\acro{CC}{convolutional coding}
\acro{Covid19}[COVID-19]{Coronavirus disease}

\acro{DAA}{detect and avoid}
\acro{DAB}{digital audio broadcasting}
\acro{DCT}{discrete cosine transform}
\acro{dft}[DFT]{discrete Fourier transform}
\acro{DR}{distortion-rate}
\acro{DS}{direct sequence}
\acro{DS-SS}{direct-sequence spread-spectrum}
\acro{DTR}{differential transmitted-reference}
\acro{DVB-H}{digital video broadcasting\,--\,handheld}
\acro{DVB-T}{digital video broadcasting\,--\,terrestrial}
\acro{DL}{DownLink}
\acro{DSSS}{Direct Sequence Spread Spectrum}
\acro{DFT-s-OFDM}{Discrete Fourier Transform-spread-Orthogonal Frequency Division Multiplexing}
\acro{DAS}{Distributed Antenna System}
\acro{DNA}{DeoxyriboNucleic Acid}

\acro{EC}{European Commission}
\acro{EED}[EED]{exact eigenvalues distribution}
\acro{EIRP}{Equivalent Isotropically Radiated Power}
\acro{ELP}{equivalent low-pass}
\acro{eMBB}{Enhanced Mobile Broadband}
\acro{EMF}{ElectroMagnetic Field}
\acro{EU}{European union}
\acro{EI}{Exposure Index}
\acro{eICIC}{enhanced Inter-Cell Interference Coordination}

\acro{FC}[FC]{fusion center}
\acro{FCC}{Federal Communications Commission}
\acro{FEC}{forward error correction}
\acro{FFT}{fast Fourier transform}
\acro{FH}{frequency-hopping}
\acro{FH-SS}{frequency-hopping spread-spectrum}
\acrodef{FS}{Frame synchronization}
\acro{FSsmall}[FS]{frame synchronization}  
\acro{FDMA}{Frequency Division Multiple Access}

\acro{GA}{Gaussian approximation}
\acro{GF}{Galois field }
\acro{GG}{Generalized-Gaussian}
\acro{GIC}[GIC]{generalized information criterion}
\acro{GLRT}{generalized likelihood ratio test}
\acro{GPS}{Global Positioning System}
\acro{GMSK}{Gaussian Minimum Shift Keying}
\acro{GSMA}{Global System for Mobile communications Association}
\acro{GS}{ground station}
\acro{GMG}{ Grid-connected MicroGeneration}

\acro{HAP}{high altitude platform}
\acro{HetNet}{Heterogeneous network}

\acro{IDR}{information distortion-rate}
\acro{IFFT}{inverse fast Fourier transform}
\acro{iht}[IHT]{iterative hard thresholding}
\acro{i.i.d.}{independent, identically distributed}
\acro{IoT}{Internet of Things}                      
\acro{IR}{impulse radio}
\acro{lric}[LRIC]{lower restricted isometry constant}
\acro{lrict}[LRICt]{lower restricted isometry constant threshold}
\acro{ISI}{intersymbol interference}
\acro{ITU}{International Telecommunication Union}
\acro{ICNIRP}{International Commission on Non-Ionizing Radiation Protection}
\acro{IEEE}{Institute of Electrical and Electronics Engineers}
\acro{ICES}{IEEE international committee on electromagnetic safety}
\acro{IEC}{International Electrotechnical Commission}
\acro{IARC}{International Agency on Research on Cancer}
\acro{IS-95}{Interim Standard 95}

\acro{KPI}{Key Performance Indicator}

\acro{LEO}{low earth orbit}
\acro{LF}{likelihood function}
\acro{LLF}{log-likelihood function}
\acro{LLR}{log-likelihood ratio}
\acro{LLRT}{log-likelihood ratio test}
\acro{LoS}{Line-of-Sight}
\acro{LRT}{likelihood ratio test}
\acro{wlric}[LWRIC]{lower weak restricted isometry constant}
\acro{wlrict}[LWRICt]{LWRIC threshold}
\acro{LPWAN}{Low Power Wide Area Network}
\acro{LoRaWAN}{Low power long Range Wide Area Network}
\acro{NLoS}{Non-Line-of-Sight}
\acro{LiFi}[Li-Fi]{light-fidelity}
 \acro{LED}{light emitting diode}
 \acro{LABS}{LoS transmission with each ABS}
 \acro{NLABS}{NLoS transmission with each ABS}

\acro{MB}{multiband}
\acro{MC}{macro cell}
\acro{MDS}{mixed distributed source}
\acro{MF}{matched filter}
\acro{m.g.f.}{moment generating function}
\acro{MI}{mutual information}
\acro{MIMO}{Multiple-Input Multiple-Output}
\acro{MISO}{multiple-input single-output}
\acrodef{maxs}[MJSO]{maximum joint support cardinality}                       
\acro{ML}[ML]{maximum likelihood}
\acro{MMSE}{minimum mean-square error}
\acro{MMV}{multiple measurement vectors}
\acrodef{MOS}{model order selection}
\acro{M-PSK}[${M}$-PSK]{$M$-ary phase shift keying}                       
\acro{M-APSK}[${M}$-PSK]{$M$-ary asymmetric PSK} 
\acro{MP}{ multi-period}
\acro{MINLP}{mixed integer non-linear programming}

\acro{M-QAM}[$M$-QAM]{$M$-ary quadrature amplitude modulation}
\acro{MRC}{maximal ratio combiner}                  
\acro{maxs}[MSO]{maximum sparsity order}                                      
\acro{M2M}{Machine-to-Machine}                                                
\acro{MUI}{multi-user interference}
\acro{mMTC}{massive Machine Type Communications}      
\acro{mm-Wave}{millimeter-wave}
\acro{MP}{mobile phone}
\acro{MPE}{maximum permissible exposure}
\acro{MAC}{media access control}
\acro{NB}{narrowband}
\acro{NBI}{narrowband interference}
\acro{NLA}{nonlinear sparse approximation}
\acro{NLOS}{Non-Line of Sight}
\acro{NTIA}{National Telecommunications and Information Administration}
\acro{NTP}{National Toxicology Program}
\acro{NHS}{National Health Service}

\acro{LOS}{Line of Sight}

\acro{OC}{optimum combining}                             
\acro{OC}{optimum combining}
\acro{ODE}{operational distortion-energy}
\acro{ODR}{operational distortion-rate}
\acro{OFDM}{Orthogonal Frequency-Division Multiplexing}
\acro{omp}[OMP]{orthogonal matching pursuit}
\acro{OSMP}[OSMP]{orthogonal subspace matching pursuit}
\acro{OQAM}{offset quadrature amplitude modulation}
\acro{OQPSK}{offset QPSK}
\acro{OFDMA}{Orthogonal Frequency-division Multiple Access}
\acro{OPEX}{Operating Expenditures}
\acro{OQPSK/PM}{OQPSK with phase modulation}

\acro{PAM}{pulse amplitude modulation}
\acro{PAR}{peak-to-average ratio}
\acrodef{pdf}[PDF]{probability density function}                      
\acro{PDF}{probability density function}
\acrodef{p.d.f.}[PDF]{probability distribution function}
\acro{PDP}{power dispersion profile}
\acro{PMF}{probability mass function}                             
\acrodef{p.m.f.}[PMF]{probability mass function}
\acro{PN}{pseudo-noise}
\acro{PPM}{pulse position modulation}
\acro{PRake}{Partial Rake}
\acro{PSD}{power spectral density}
\acro{PSEP}{pairwise synchronization error probability}
\acro{PSK}{phase shift keying}
\acro{PD}{power density}
\acro{8-PSK}[$8$-PSK]{$8$-phase shift keying}
\acro{PPP}{Poisson point process}
\acro{PCP}{Poisson cluster process}
 
\acro{FSK}{Frequency Shift Keying}

\acro{QAM}{Quadrature Amplitude Modulation}
\acro{QPSK}{Quadrature Phase Shift Keying}
\acro{OQPSK/PM}{OQPSK with phase modulator }

\acro{RD}[RD]{raw data}
\acro{RDL}{"random data limit"}
\acro{ric}[RIC]{restricted isometry constant}
\acro{rict}[RICt]{restricted isometry constant threshold}
\acro{rip}[RIP]{restricted isometry property}
\acro{ROC}{receiver operating characteristic}
\acro{rq}[RQ]{Raleigh quotient}
\acro{RS}[RS]{Reed-Solomon}
\acro{RSC}[RSSC]{RS based source coding}
\acro{r.v.}{random variable}                               
\acro{R.V.}{random vector}
\acro{RMS}{root mean square}
\acro{RFR}{radiofrequency radiation}
\acro{RIS}{Reconfigurable Intelligent Surface}
\acro{RNA}{RiboNucleic Acid}
\acro{RRM}{Radio Resource Management}
\acro{RUE}{reference user equipments}
\acro{RAT}{radio access technology}
\acro{RB}{resource block}

\acro{SA}[SA-Music]{subspace-augmented MUSIC with OSMP}
\acro{SC}{small cell}
\acro{SCBSES}[SCBSES]{Source Compression Based Syndrome Encoding Scheme}
\acro{SCM}{sample covariance matrix}
\acro{SEP}{symbol error probability}
\acro{SG}[SG]{sparse-land Gaussian model}
\acro{SIMO}{single-input multiple-output}
\acro{SINR}{signal-to-interference plus noise ratio}
\acro{SIR}{signal-to-interference ratio}
\acro{SISO}{Single-Input Single-Output}
\acro{SMV}{single measurement vector}
\acro{SNR}[\textrm{SNR}]{signal-to-noise ratio} 
\acro{sp}[SP]{subspace pursuit}
\acro{SS}{spread spectrum}
\acro{SW}{sync word}
\acro{SAR}{specific absorption rate}
\acro{SSB}{synchronization signal block}
\acro{SR}{shrink and realign}

\acro{tUAV}{tethered Unmanned Aerial Vehicle}
\acro{TBS}{terrestrial base station}

\acro{uUAV}{untethered Unmanned Aerial Vehicle}
\acro{PDF}{probability density functions}

\acro{PL}{path-loss}

\acro{TH}{time-hopping}
\acro{ToA}{time-of-arrival}
\acro{TR}{transmitted-reference}
\acro{TW}{Tracy-Widom}
\acro{TWDT}{TW Distribution Tail}
\acro{TCM}{trellis coded modulation}
\acro{TDD}{Time-Division Duplexing}
\acro{TDMA}{Time Division Multiple Access}
\acro{Tx}{average transmit}

\acro{UAV}{Unmanned Aerial Vehicle}
\acro{uric}[URIC]{upper restricted isometry constant}
\acro{urict}[URICt]{upper restricted isometry constant threshold}
\acro{UWB}{ultrawide band}
\acro{UWBcap}[UWB]{Ultrawide band}   
\acro{URLLC}{Ultra Reliable Low Latency Communications}
         
\acro{wuric}[UWRIC]{upper weak restricted isometry constant}
\acro{wurict}[UWRICt]{UWRIC threshold}                
\acro{UE}{User Equipment}
\acro{UL}{UpLink}

\acro{WiM}[WiM]{weigh-in-motion}
\acro{WLAN}{wireless local area network}
\acro{wm}[WM]{Wishart matrix}                               
\acroplural{wm}[WM]{Wishart matrices}
\acro{WMAN}{wireless metropolitan area network}
\acro{WPAN}{wireless personal area network}
\acro{wric}[WRIC]{weak restricted isometry constant}
\acro{wrict}[WRICt]{weak restricted isometry constant thresholds}
\acro{wrip}[WRIP]{weak restricted isometry property}
\acro{WSN}{wireless sensor network}                        
\acro{WSS}{Wide-Sense Stationary}
\acro{WHO}{World Health Organization}
\acro{Wi-Fi}{Wireless Fidelity}

\acro{sss}[SpaSoSEnc]{sparse source syndrome encoding}

\acro{VLC}{Visible Light Communication}
\acro{VPN}{Virtual Private Network} 
\acro{RF}{Radio Frequency}
\acro{FSO}{Free Space Optics}
\acro{IoST}{Internet of Space Things}

\acro{GSM}{Global System for Mobile Communications}
\acro{2G}{Second-generation cellular network}
\acro{3G}{Third-generation cellular network}
\acro{4G}{Fourth-generation cellular network}
\acro{5G}{Fifth-generation cellular network}	
\acro{gNB}{next-generation Node-B Base Station}
\acro{NR}{New Radio}
\acro{UMTS}{Universal Mobile Telecommunications Service}
\acro{LTE}{Long Term Evolution}

\acro{QoS}{Quality of Service}
\end{acronym}
	
\newcommand{\SAR} {\mathrm{SAR}}
\newcommand{\WBSAR} {\mathrm{SAR}_{\mathsf{WB}}}
\newcommand{\gSAR} {\mathrm{SAR}_{10\si{\gram}}}
\newcommand{\Sab} {S_{\mathsf{ab}}}
\newcommand{\Eavg} {E_{\mathsf{avg}}}
\newcommand{\ft}{f_{\textsf{th}}}
\newcommand{\alphatf}{\alpha_{24}}

\title{Non-Terrestrial Network Models Using Stochastic Geometry: Planar or Spherical?}
\author{
Ruibo~Wang,~\IEEEmembership{Member,~IEEE,} Baha~Eddine~Youcef~Belmekki,~\IEEEmembership{Senior Member,~IEEE,} \\
Howard~H.~Yang,~\IEEEmembership{Member,~IEEE,} and Mohamed~Slim~Alouini,~\IEEEmembership{Fellow,~IEEE}
\thanks{Ruibo Wang and Mohamed-Slim Alouini are with King Abdullah University of Science and Technology (KAUST), CEMSE division, Thuwal 23955-6900, Saudi Arabia. Baha Eddine Youcef Belmekki is with the School of Engineering and Physical Sciences, Heriot-Watt University, Edinburgh EH14 4AS, United Kingdom. Howard H. Yang is with the ZJU-UIUC Institute, Zhejiang University, Haining 314400, China. (e-mail: ruibo.wang@kaust.edu.sa; b.belmekki@hw.ac.uk; haoyang@intl.zju.edu.cn; slim.alouini@kaust.edu.sa). Corresponding author: Howard H. Yang. 
}
\vspace{-8mm}
}
\maketitle

\vspace{-0.8cm}

{\color{black}
\begin{abstract}
With the explosive deployment of non-terrestrial networks (NTNs), the computational complexity of network performance analysis is rapidly escalating. As one of the most suitable mathematical tools for analyzing large-scale network topologies, stochastic geometry (SG) enables the representation of network performance metrics as functions of network parameters, thus offering low-complexity performance analysis solutions. However, choosing between planar and spherical models remains challenging. Planar models neglect Earth's curvature, causing deviations in high-altitude NTN analysis, yet are still often used for simplicity. This paper introduces relative error to quantify the gap between planar and spherical models, helping determine when planar modeling is sufficient. To calculate the relative error, we first propose a point process (PP) generation algorithm that simultaneously generates a pair of homogeneous and asymptotically similar planar and spherical PPs. We then introduce several typical similarity metrics, including topology-related and network-level metrics, and further develop a relative error estimation algorithm based on these metrics. In addition, we derive an analytical expression for the optimal planar altitude, which reduces computational complexity and provides theoretical support for planar approximation. Finally, numerical results investigate how deployment altitude and region affect NTN modeling, with case studies on HAP and LEO satellite constellations.
\end{abstract}
}


\section{Introduction} \label{section1}
{\color{black}Non-terrestrial networks (NTNs) are pivotal in wireless communication networks due to their irreplaceable advantages in enhancing coverage, ensuring stable connectivity, and guaranteeing seamless communications \cite{belmekki2022unleashing,belmekki2024cellular, he2024direct}. However, the expanding scale of non-terrestrial platforms (NTPs), and their non-stationary inherent and coherent interference in mega networks significantly increase the computational complexity of performance evaluation. To address this issue, stochastic geometry (SG) utilizes specific random point process distributions to model satellite positions, enabling powerful analytical capabilities. With this capability, we can derive analytical expressions for key performance metrics (e.g., data rate), allowing for an exact mapping from input parameters (e.g., satellite altitude) to performance metrics. Since this mapping is deterministic, the analytical approach significantly reduces computational complexity \cite{wang2025modeling}. Compared to the current state-of-the-art numerical simulation methods, the computational delay of this framework is proved to be only one four-thousandth of that of the simulation \cite{qiu2023performance}. Meanwhile, SG is one of the most suitable mathematical tools for modeling massive dynamic topologies and possesses unique interference analysis capabilities \cite{wang2022ultra}.
}

\par
However, there is ongoing debate over whether to use a planar or spherical model for NTN modeling. Due to the significantly higher complexity of the spherical model compared to the planar model, some researchers, considering the need for low-complexity performance analysis results, prefer to use planar modeling \cite{gao2019spectrum}. Another group of researchers holds the opposite view. They consider that SG-based models are already approximations, as they assume NTPs follow a specific random distribution to achieve analytical traceability \cite{alzenad2019coverage}. If the deployment area is further approximated, the model, having undergone two levels of approximation, may no longer be accurate.

\par
To balance between analytical simplicity and modeling accuracy, researchers have collectively agreed upon the following guidelines for NTN modeling. When the altitude of NTPs is below $20$~km, researchers generally believe that the curvature of the Earth has a negligible impact on modeling, making spherical modeling unnecessary. For instance, authors model unmanned aerial vehicles (UAVs) at an altitude of $5$~km \cite{zhang2021stochastic} and aircraft at an altitude of $10$~km \cite{tian2023satellite} as planar point processes (PPs). 
Conversely, for NTPs at altitudes above $20$~km, spherical modeling is more appropriate \cite{okati2022nonhomogeneous}. It is unreasonable to approximate a constellation of satellites distributed on a sphere as a planar model. Using altitude to determine whether to use planar or spherical modeling is an interesting criterion. However, this criterion is empirical rather than quantitative, making its limitations and potential shortcomings apparent.

\par
Firstly, when planar modeling is applied to devices slightly below $20$~km and slightly above $20$~km, the modeling errors may be similar. Therefore, using $20$~km as an artificial dividing line lacks a theoretical basis. Secondly, there is debate over whether a HAP network at an altitude of $20$~km should be modeled as planar or spherical. Some researchers believe that planar modeling is more straightforward \cite{matracia2023uav}, while others consider that spherical modeling is more reasonable \cite{232306,232304,242302}. 
Notably, the authors in \cite{gao2019spectrum} considered spherical modeling to be more reasonable for HAPs. However, due to the excessive complexity of the analytical process, they eventually mapped the spherical model back to a planar one. 

\par
Finally, in addition to altitude, other factors, such as the area of the distribution region, are key determinants of the error introduced by planar approximation. The author in \cite{sun2024performance} modeled the users within the service area of a single satellite as a planar PP. In this work, the satellite only serves users within the main lobe width of the directional beam, thus the service area is relatively small. In addition, although the author in \cite{hu2024performance} modeled the ground equipment as a spherical PP, they still approximated the ground equipment within the satellite's service area as a planar distribution during the topological analysis. Conversely, if users are equipped with directional beams for uplink transmission, approximating the satellites within the user's beam range as a planar distribution is also somewhat reasonable.

\par
Based on the above analysis, it is evident that the issue of whether to use planar or spherical modeling in SG-based research has sparked considerable debate. The reason for the controversy is that the choice between the two modeling approaches involves many factors, and relying solely on qualitative experience is not rigorous. Therefore, it is necessary to quantitatively analyze the suitability of using planar or spherical modeling for NTNs and to investigate the impact of various factors on the choice of modeling through case studies. Sec.~\ref{section3}, Sec.~\ref{section4}, and Sec.~\ref{section5} sequentially present the process of the aforementioned quantitative analysis. Sec.~\ref{section2}, however, presents the research approach in reverse order to highlight the motivation behind the case studies and algorithm design.

\section{Research Structure} \label{section2}
This section presents the three objectives of this article and explains the logical connections between them. Furthermore, we provide the relevant literature foundation and contributions corresponding to each of the three objectives.

\subsection{Objective} \label{section2-1}
{\color{black}
This article has three objectives: similarity measure, error estimation algorithm design, and PP generation algorithm design. The details are as follows.
\begin{itemize}
    \item \textbf{Objective 1 (Similarity Measure)}:     Considering that when the difference between the planar and spherical models is sufficiently small, adopting the planar model to reduce analytical complexity is a better option. Therefore, we can quantitatively express the necessity of spherical modeling, which is a qualitative issue, through similarity measures.
    \item \textbf{Objective 2 (Error Estimation)}: For ease of analysis, the relative error is introduced to measure the similarity between two types of models. First, we map the planar model and the spherical model to their corresponding performance metrics, such as coverage probability. Next, the relative error is calculated by the ratio of the absolute difference between the two metrics to the spherical model's metric. The above process is implemented through the error estimation algorithm.
    \item \textbf{Objective 3 (PP Generation)}: To estimate the relative error, we need to generate paired planar and spherical PPs. The generated PPs should be similar in both spatial distribution and statistical characteristics. This is not only to address the need for most similar approximations but also to obtain deterministic performance metrics and relative error, ensuring that the similarity measure results between PPs are not random. For this purpose, it is necessary to design a PP generation algorithm.
\end{itemize}
}

\par
The above analysis shows that each subsequent objective lays the foundation for the previous one. The necessity of spherical modeling and similarity measurement requires the implementation of error estimation algorithms. The estimation of relative error, in turn, is based on pairs of similar planar and spherical PPs generated by the PP generation algorithm.

\subsection{Related Works}
This subsection provides the works related to the three objectives. Regarding Objective 1, related works are limited to qualitatively analyzing the choice between spherical and planar modeling. These discussions have already been covered in Sec.~\ref{section1}. This is the first study that quantitatively analyzes this issue by introducing similarity measures and relative error. 

\par
For Objective 2, some literature has considered measuring the accuracy of SG-based NTN models through error estimation. As mentioned, the SG framework assumes that NTPs follow a specific stochastic distribution to achieve analytical tractability. Given that satellites in actual constellations are deterministically distributed based on fixed configurations, existing studies have comprehensively demonstrated the accuracy of using spherical SG models in performance analysis. These studies thoroughly cover non-orbital models \cite{yastrebova2020theoretical,al2021tractable,okati2020downlink}, stochastic orbital models \cite{choi2024modeling}, and fixed orbital models \cite{jung2023modeling,wang2022evaluating}. The performance metrics used for error estimation include both topological-related metrics \cite{al2021session,jung2023modeling,wang2022evaluating} and system-level metrics \cite{yastrebova2020theoretical,okati2020downlink,choi2024modeling}. The comparisons also involve simulation software like Systems Tool Kit (STK) \cite{yastrebova2020theoretical} and deterministic constellation configurations, such as Walker constellations \cite{okati2020downlink,al2021tractable}. These studies first map PPs or actual constellations to performance metrics through Monte Carlo simulations, and then estimate errors based on differences between these metrics. Therefore, they have inspired this paper's research approach. However, since our research contents are different from those studies, there are significant differences in technical methods, case study designs, and analytical conclusions. For instance, in terms of technical methods, none of the above studies involve the paired PP generation algorithm.

\par
To simplify derivations, some research has implemented a mapping from a spherical PP to a planar one, which is similar to Objective 3. In \cite{hu2024performance}, the authors considered the communication between ships to a typical user relayed by a satellite. They projected the ships onto the horizontal surface of the typical user, keeping the distance between them and the user unchanged. However, this projection method results in an underestimation of the distances from all ships to the satellite. In contrast, the authors in \cite{gao2019spectrum} achieved a planar mapping by converting the coordinates of HAPs from spherical coordinates to cylindrical coordinates, resulting in an underestimation of the distances from all HAPs to the typical ground user. The planar models obtained from these two mapping methods are not only dissimilar to the original spherical models but also lose the statistical characteristics of the spherical models, such as homogeneity. Although the mapped planar PPs are no longer homogeneous, the authors in \cite{hu2024performance} and \cite{gao2019spectrum} still used properties of homogeneous PP to simplify derivations, which constitutes a technical flaw. Therefore, we need to redesign a PP generation algorithm that ensures the planar PP has the same statistical characteristics as the spherical PP while being as similar as possible. {\color{black} It is worth mentioning that recent research has proposed an area-preserving mapping strategy that maps planar PP to spherical PP, providing a novel approach to the mapping problem \cite{angervuori2025meta}.}

\subsection{Contribution}
Corresponding to the objectives in Sec.~\ref{section2-1}, the contributions of this article are as follows:
\begin{itemize}
    \item \textbf{Contribution 1 (PP Generation)}: We propose a PP generation algorithm that produces a pair of planar and spherical PPs, and we demonstrate that both the generated PPs are homogeneous and asymptotically similar. Additionally, by deriving the contact distance distribution from the NTP to the typical user, we show that the analytical expression of the spherical PP is more complex than that of the planar PP.
    \item \textbf{Contribution 2 (Error Estimation)}: We design an error estimation algorithm to evaluate the relative error between a pair of planar and spherical PPs. The computational complexity of the error estimation algorithm is analyzed. Six performance metrics are introduced to calculate the relative error: three are topology-related metrics, and three are system-level metrics. In addition, we derive the analytical result for the optimal altitude of the planar PP that minimizes the relative error with the spherical PP. 
    \item \textbf{Contribution 3 (Similarity Measure)}: We design three case studies to discuss the impact of different performance metrics, the altitude of the NTN, and the half-power beamwidth on the relative error. Additionally, the case studies demonstrate how the quantified relative error can guide the choice between spherical and planar modeling.
\end{itemize}

\par
Based on the above analysis, the core contributions of this paper are to provide recommendations for NTN modeling and to offer theoretical support for planar approximations within a limited range. It thus significantly differs from traditional SG works, which aim at establishing traceable performance analysis frameworks.

\section{PP Generation} \label{section3}
This section focuses on the design of the PP generation algorithm. In addition to the algorithm design, we present two criteria for PP generation and derive some analytical results based on the PPs generated by the algorithm.

\subsection{PP Generation Criteria} \label{section3-1}
{\color{black}The first criterion that we expect the generated point process to satisfy is homogeneity.} Homogeneity is one of the most common assumptions in spherical SG, and most of the studies assume that the NTPs follow a homogeneous PP. The mathematical definition of homogeneity is as follows. 

\begin{definition}[Homogeneity]
When the number of NTPs within any region of a PP is proportional to the area of the region, the PP is said to be homogeneous.
\end{definition}

When the NTN is modeled as a homogeneous PP, the NTPs follow the same distribution from the perspective of any user, regardless of the user's location. As a result, the analytical expressions for performance metrics under a homogeneous PP model are user's coordinates-independent, simplifying the analysis.

\par
Next, we explain why two PPs need to be similar and provide the motivation for introducing asymptotic similarity. Assume that a deterministic satellite constellation at an altitude of $550$~km can provide coverage with an $80\%$ probability for a typical user. Existing studies have proven that modeling the satellites as a random PP on the $550$~km spherical surface, or as a PP located on the spherical cap corresponding to the user's line-of-sight, can accurately estimate the $80\%$ coverage probability. However, to ensure that the planar PP model can still estimate a coverage probability close to $80\%$, the PP should be located directly above the user with an altitude a little lower than $550$~km. If the planar PP deviates from this region or is positioned below the horizon, the estimated coverage probability has a large discrepancy from the actual result. This example illustrates the necessity of similarity in spatial distribution. The asymptotic similarity proposed in this paper imposes even stricter requirements on the similarity between two PPs.

\begin{definition}[Asymptotic Similarity]
Given that the spherical PP is distributed over a spherical cap region with a finite area of $S$ (which can be arbitrary). When the radius of the spherical cap $R_s$ tends to infinity, if the spherical PP coincides with the planar PP, these two PPs are asymptotically similar. 
\end{definition}

\par
\begin{figure}[ht]
\centering
\includegraphics[width=\linewidth]{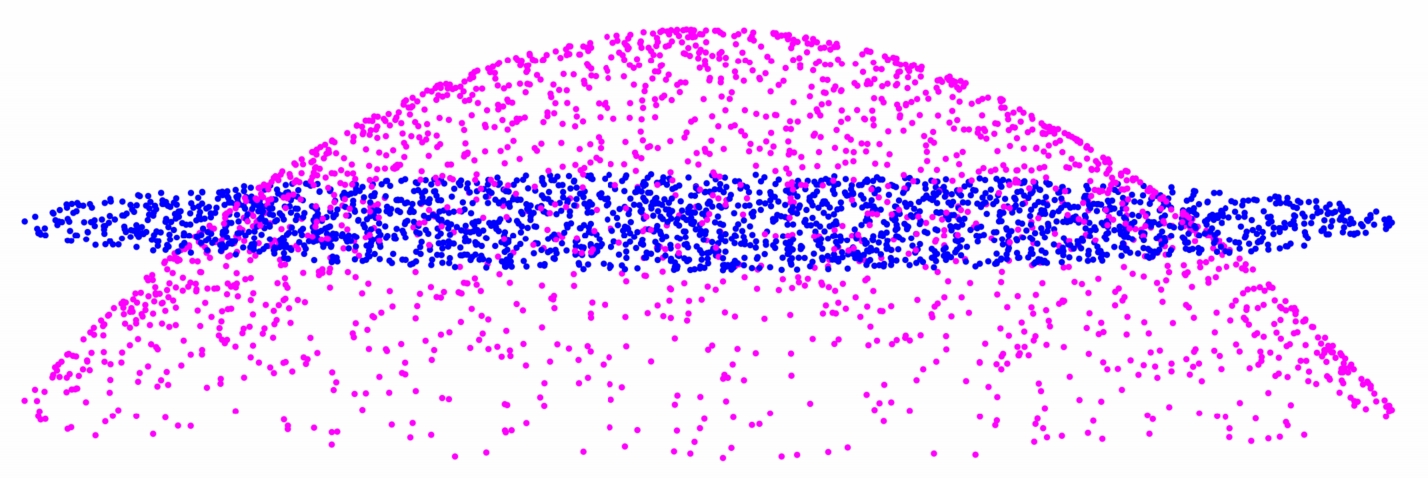}
\caption{Result diagram implemented by Algorithm~\ref{alg1}.}
\label{figure1}
\end{figure}

Next, we use Fig.~\ref{figure1} to explain the meaning of asymptotic similarity. Given that the area $S$ of the spherical cap is fixed. As the radius $R_s$ increases, the curvature of the spherical cap gradually decreases, and the spherical cap eventually becomes a plane. In addition to requiring spatial distribution similarity, asymptotic similarity demands that the coordinates of each point in the spherical PP coincide with the coordinates of one of the points in the planar PP, as $R_s$ approaches infinity.

\subsection{PP Generation Algorithm}
Before describing the algorithm, the distribution region is determined. We set the center of the Earth as the origin and observe the generated PPs from the perspective of a typical ground user. Since the PPs are required to be homogeneous, the distributions of the PPs remain unchanged with the rotation of the coordinate axes, according to Slivnyak's theorem \cite{feller1991introduction}. Without loss of generality, we assume the coordinates of the typical user in the spherical coordinate system are $(R_\oplus, 0, 0)$, where $R_\oplus$ is the radius of the Earth. Considering that NTPs below the user's horizon are blocked by the Earth and cannot communicate with the user via direct links, we only model the NTN above the horizon. The region where these NTPs are deployed forms a finite spherical cap, and their distribution is therefore assumed to follow a binomial point process (BPP). In addition to being suitable for modeling finite regions, BPP is also one of the most widely applied types of point processes in spherical SG \cite{wang2024ultra}.

\par
The rotation axis for PPs is the line passing through the Earth center and the user. To facilitate the expression, the coordinates of an NTP from the spherical PP are denoted as $x(R_s,\varphi_s,\theta_s)$, where $R_s$, $\varphi_s$ in the spherical coordinate system, where $R_s$, $\varphi_s$, and $\theta_s$ represent the radius, azimuth angle, and polar angle, respectively. The coordinates of an NTP from the planar PP are denoted as $y(\rho_p,\varphi_p,h_p)$, where $\rho_p$, $\varphi_p$, and $h_p$ represent the horizontal radius, azimuth angle, and altitude, respectively. The PP generation algorithm is proposed as follows.

\begin{algorithm}[!ht] 
    \caption{PP Generation $(\mathcal{X},\mathcal{Y}) \leftarrow \mathcal{F}(R_s,\theta_{\max}; \rho_{\max},h_p)$}
    \label{alg1} 
    \begin{algorithmic} [1]
    \STATE \textbf{Input}: Radius $R_s$, maximum polar angle of the spherical cap $\theta_{\max}$, maximum horizontal radius $\rho_{\max}$ and altitude $h_p$.

    \FOR{$n = 1 : N_p$}
    \STATE $u^{(n)} \leftarrow {\mathrm{rand}}(0,1)$, $v^{(n)} \leftarrow {\mathrm{rand}}(0,1)$.
    \STATE $\theta_s^{(n)} \leftarrow \arccos(1-u^{(n)} (1-\cos\theta_{\max}))$, \\ $\rho_p^{(n)}\leftarrow \sqrt{u^{(n)}} \rho_{\max}$.
    \STATE $\varphi_s^{(n)} \leftarrow 2\pi v^{(n)}$, $\varphi_p^{(n)} \leftarrow 2\pi v^{(n)}$.
    \ENDFOR
    \STATE \textbf{Output}: Spherical PP $\mathcal{X} \leftarrow \Big\{ x^{(1)} \left(R_s,\varphi_s^{(1)},\theta_s^{(1)} \right),$ \\ $x^{(2)}\left(R_s, \varphi_s^{(2)} ,\theta_s^{(2)}\right), \dots, x^{(N_p)} \left(R_s, \varphi_s^{(N_p)}, \theta_s^{(N_p)}\right) \Big\}$, and planar PP $\mathcal{Y} \leftarrow \Big\{ y^{(1)} \left(\rho_p^{(1)},\varphi_p^{(1)},h_p \right),$ \\ $y^{(2)}\left(\rho_p^{(2)}, \varphi_p^{(2)} ,h_p\right), \dots,y^{(N_p)} \left(\rho_p^{(N_p)}, \varphi_p^{(N_p)}, h_p\right) \Big\}.$
    \end{algorithmic}
\end{algorithm}	
In step (2) of Algorithm~\ref{alg1}, $N_p$ is the number of NTPs in the spherical PP or planar PP. In step (3), ${\mathrm{rand}}(0,1)$ refers to generating a number between $0$ and $1$ uniformly at random. The operator $\mathcal{F}$ represents the mapping from distribution parameters $(R_s,\theta_{\max}; \rho_{\max},h_p)$ to two PPs $(\mathcal{X},\mathcal{Y})$ by Algorithm~\ref{alg1}.

\subsection{Analytical Results}
{\color{black}The following theorems prove that both PPs obtained by Algorithm~\ref{alg1} are homogeneous, and provide sufficient conditions for asymptotic similarity.}

\begin{theorem}\label{theorem1}
    The spherical BPP $\mathcal{X}$ and the planar BPP $\mathcal{Y}$ are both homogeneous.
\end{theorem}
\begin{IEEEproof}
See Appendix~\ref{app:theorem1}.
\end{IEEEproof}

\par

\begin{theorem}\label{theorem2}
When $\rho_{\max} = R_s \sin\theta_{\max}$ and $R_s \cos\theta_{\max} < h_p < R_s$ are satisfied, the spherical BPP $\mathcal{X}$ and the planar BPP $\mathcal{Y}$ are asymptotically similar.
\end{theorem}
\begin{IEEEproof}
See Appendix~\ref{app:theorem2}.
\end{IEEEproof}

\par
The conditions in Theorem~\ref{theorem2} are required to be satisfied in the subsequent parts. Based on the above theorems, the planar PP mapped from the spherical PP in the existing literature is neither homogeneous nor asymptotically similar to the spherical one \cite{hu2024performance,gao2019spectrum}. Therefore, Algorithm~\ref{alg1} not only lays the foundation for the design of error estimation algorithms but also provides a mapping scheme. Hereafter, we derive the distance distribution, which is one of the most basic analytical results in the SG framework.

\begin{lemma}\label{lemma1}
The cumulative distribution function (CDF) of the distance distribution from the user to the NTP $x^{(n)}$ is
\begin{equation}
\begin{split}
& F_s (d) = \frac{1}{1-\cos\theta_{\max}} \left( 1 - \frac{R_{\oplus}^2 + R_s^2 - d^2}{2 R_{\oplus} R_s} \right), \\
& \ \ \ \ \ R_s - R_{\oplus} \leq d \leq \sqrt{R_{\oplus}^2 + R_s^2 - 2R_{\oplus} R_s \cos\theta_{\max}},
\end{split}
\end{equation}
where $\theta_{\max}$ denotes the maximum polar angle of the spherical cap. The CDF of the distance distribution from the user to the NTP $y^{(n)}$ is
\begin{equation}
    F_p (d) = \frac{d^2-h_p^2}{\rho_{\max}^2}, \ \ h_p \leq d \leq \sqrt{h_p^2 + \rho_{\max}^2},
\end{equation}
where $\rho_{\max}$ represents the maximum horizontal radius.
\begin{IEEEproof}
See Appendix~\ref{app:lemma1}.
\end{IEEEproof}
\end{lemma}

The lemma clearly illustrates that planar modeling is more convenient for analysis compared to spherical modeling. For example, the impact of altitude $h_p$ on the distance distribution $F_p (d)$ is easier to observe intuitively under the planar PP model, while the influence of $R_s$ (which is related to the altitude) on the distance distribution $F_s (d)$ under the spherical PP model is less straightforward.

\section{Error Estimation} \label{section4}
This section develops an algorithm to measure the relative error between $\mathcal{X}$ and $\mathcal{Y}$ generated by Algorithm~\ref{alg1}, which applies to any performance metric. Furthermore, we provide six representative performance metrics. To broaden the applicability and reduce computational complexity, we also derive the optimal altitude for the planar PP.

\subsection{Error Estimation Algorithm} \label{section4-1}
In this section, we present the implementation of the error estimation algorithm, which is Algorithm~\ref{alg2}. The design of the error estimation algorithm involves two challenges: defining the relative error and determining the optimal altitude. The former is addressed in steps (7) and (8) of Algorithm~\ref{alg2}, and will be explained in detail in Sec.~\ref{section4-2}. The latter corresponds to the outer loop of Algorithm~\ref{alg2} and is discussed in Sec.~\ref{section4-3}.

\begin{algorithm}[!ht] 
    \caption{Error Estimation Algorithm}
    \label{alg2} 
    \begin{algorithmic} [1]
    \STATE \textbf{Input}: Radius $R_s$ and maximum polar angle of the spherical cap $\theta_{\max}$.

    \STATE \textbf{Initialize}: $E_{\min} \leftarrow 1$, $\Delta h \leftarrow \frac{R_s - R_s \cos\theta_{\max}}{N_{\mathrm{out}}+1}$, and $h_p \leftarrow R_s \cos\theta_{\max}$.
        
    \FOR{$n_{\mathrm{out}} = 1 : N_{\mathrm{out}}$}
    \STATE $\Delta E \leftarrow 0$, $h_p \leftarrow h_p + \Delta h$.

    \FOR{$n_{\mathrm{in}} = 1 : N_{\mathrm{in}}$}
    \STATE $(\mathcal{X},\mathcal{Y}) \leftarrow \mathcal{F}(R_s,\theta_{\max}; \rho_{\max},h_p)$.
    \STATE $g_s \leftarrow \mathcal{G}(\mathcal{X})$, $g_p \leftarrow \mathcal{G}(\mathcal{Y})$.
    \STATE $\Delta E \leftarrow \frac{|g_s - g_p|}{g_s} + \Delta E$.
        
    \ENDFOR

    \IF{$E_{\min} > \frac{\Delta E}{N_{\mathrm{in}}}$}
    \STATE $E_{\min} \leftarrow \frac{\Delta E}{N_{\mathrm{in}}}$, $h_{\mathrm{opt}} \leftarrow h_p$.
    \ENDIF
        
    \ENDFOR
    \STATE \textbf{Output}: Relative error $E_{\min}$ and the corresponding optimal altitude $h_{\mathrm{opt}}$.  
    \end{algorithmic}
\end{algorithm}	

In Algorithm~\ref{alg2}, $N_{\mathrm{in}}$ and $N_{\mathrm{out}}$ are numbers of iterations for the inner and outer loops, respectively, $|\cdot|$ denotes the absolute value function.

\subsection{Performance Metrics} \label{section4-2}
Algorithm~\ref{alg2} provides a general framework applicable to all performance metrics. In step (7), the function $\mathcal{G}(\mathcal{X})$ maps the PP $\mathcal{X}$ to a metric. This subsection presents six representative metrics, most of which have already been applied to error evaluation between SG-based models and deterministic satellite constellations. The three topological-related metrics are:

\begin{itemize}
    \item $(t_1)$: Minimum energy required to move from $\mathcal{X}$ to $\mathcal{Y}$. 
    \item $(t_2)$: Minimum average energy required to move $x^{(n)}$ or $y^{(n)}$ to the typical user located at $(R_{\oplus}, 0, 0)$.
    \item $(t_3)$: Energy required to move the point closest to the typical user in $\mathcal{X}$ or $\mathcal{Y}$ to $(R_{\oplus}, 0, 0)$.
\end{itemize}

\par
$(t_1)$ is known as the Earth mover's distance \cite{rubner1998metric}. $(t_2)$ is the Wasserstein distance in the discrete case and has been applied for error estimation under the SG framework in \cite{wang2022evaluating}. $(t_3)$ represents the contact distance proposed in \cite{talgat2020nearest} and is used for error estimation in \cite{jung2023modeling}.

\par
There are still some aspects regarding topology-related metrics that need to be clarified. Without loss of generality, we assume that each point has a unit mass in these topological-related metrics. $(t_1)$ can be calculated by changing steps (7) and (8) in Algorithm~\ref{alg2} into $\Delta E \leftarrow \mathcal{G}(\mathcal{X},\mathcal{Y}) + \Delta E$, where $\mathcal{G}(\mathcal{X},\mathcal{Y})$ represents the average energy required to move $x^{(n)}$ to $y^{(n)}$ divided by the energy required to move $x^{(n)}$ to the origin $n$ can be any positive integer less than or equal to $N_p$. Note that in order to compare $(t_1)$ with $(t_2)$ and $(t_3)$ on the same scale, we have scaled $(t_1)$ by dividing it by the energy to move $x^{(n)}$ to the origin.

\par
Additionally, we select the widely studied system-level performance metrics under the SG framework:
\begin{itemize}
    \item $(s_1)$: Average signal-to-interference plus noise ratio (SINR).
    \item $(s_2)$: Coverage probability.
    \item $(s_3)$: Average achievable data rate.
\end{itemize}

\par
$(s_2)$ is the probability that the received SINR is larger than a threshold, and $(s_3)$ is the ergodic capacity from the Shannon-Hartley theorem over
a fading communication link. $(s_2)$ and $(s_3)$ are used in error evaluation between SG-based models and deterministic constellations in \cite{choi2024modeling} and \cite{okati2020downlink}, respectively. Considering that performance metrics other than coverage probability are not normalized, we define $\frac{|\mathcal{G}(\mathcal{X}) - \mathcal{G}(\mathcal{Y})|}{\mathcal{G}(\mathcal{X})} $ as the relative error in step (8) of Algorithm~\ref{alg2}.

\subsection{Optimal Altitude and Computational Complexity} \label{section4-3}
Theorem~\ref{theorem2} only provides the altitude range under asymptotic optimality, whereas the relative error varies at different altitudes. Therefore, the Algorithm~\ref{alg2} traverses the altitudes within this range through an outer loop to find the optimal altitude $h_{\mathrm{opt}}$ where the relative error is minimized, ensuring the uniqueness of the relative error obtained by the algorithm. However, the method in Algorithm~\ref{alg2} has two drawbacks. First, adding an extra loop results in higher computational complexity. Second, the optimal altitude varies with changes in performance metrics, limiting the scalability of the algorithm.

\par
Therefore, we use $(t_2)$ as the performance metric to derive the optimal altitude $h_{\mathrm{opt}}$ that minimizes the relative error. This altitude can serve as a recommended reference for performing the spherical-to-planar mapping. Before deriving $h_{\mathrm{opt}}$, we assume $N_s \rightarrow \infty$ as a premise to achieve the equivalent effect of the inner loop in Algorithm~\ref{alg2}. As a result, $h_{\mathrm{opt}}$ can be obtained by solving the following optimization problem when $N_s \rightarrow \infty$:
\begin{equation}
\begin{split}
    \mathscr{P}: h_{\mathrm{opt}} = \underset{h_p}{\argmin} \frac{\left| \mathcal{G}(\mathcal{X}) - \mathcal{G}(\mathcal{Y}) \right|}{\mathcal{G}(\mathcal{X})},
\end{split}
\end{equation}
where $\mathcal{G}(\mathcal{X})$ maps the PP $\mathcal{X}$ to the metric $(t_2)$.

\begin{proposition}\label{prop1}
The solution of problem $\mathscr{P}$ is given as
\begin{equation}
\begin{split}
    & h_{\mathrm{opt}} = R_{\oplus} \\
    & + \sqrt{R_{\oplus}^2 - \frac{1}{2}\rho_{\max}^2 - (1+\cos\theta_{\max}) R_s R_{\oplus} + R_s^2}.
\end{split}
\end{equation}
\end{proposition}
\begin{IEEEproof}
See Appendix~\ref{app:prop1}.
\end{IEEEproof}

\par
Note that the numerical results in Sec.~\ref{section5} will demonstrate that the choice of metrics has little impact on the optimal altitude, thereby validating that the optimal altitude provided in Proposition~\ref{prop1} can be generalized to other metrics.

\par
Based on Proposition~\ref{prop1}, the computational complexity of the algorithm can be analyzed. The mapping of $\mathcal{G}(\mathcal{X})$ and $\mathcal{G}(\mathcal{Y})$ constitutes a significant part of the computational complexity compared to other steps. Hence, we consider one round execution of $\mathcal{G}(\mathcal{X})$ or $\mathcal{G}(\mathcal{Y})$ as the unit complexity. The computational complexity of relative error depends on which of the following three methods is adopted.
\begin{itemize}
    \item Complete simulation method (CSM): When Algorithm~\ref{alg2} is fully executed, the computational complexity is $\mathcal{O}(N_{\mathrm{in}}N_{\mathrm{out}})$.
    \item Joint simulation and analysis method (JSAM): Replace the outer loop of Algorithm~\ref{alg2} with the recommended value of $h_{\mathrm{opt}}$ provided in Proposition~\ref{prop1}. In this case, the computational complexity is $\mathcal{O}(N_{\mathrm{in}})$.
    \item Complete analysis method (CAM): Most of the SG-based studies can achieve a direct mapping from parameters such as $R_s$ and $\theta_{\max}$ to feature metrics, and the computational complexity is only $\mathcal{O}(1)$.
\end{itemize}

\par
Based on the above analysis, CAM has the lowest computational complexity. However, executing CAM requires the analytical results of metrics under both types of PP modeling, and currently, no existing literature has achieved this. In contrast, CSM and JSAM can be executed without extra derivation, and thus are adopted in the next section. Compared to CSM, JSAM has lower computational complexity but provides approximate results. For the sake of accuracy, unless otherwise specified, the numerical results in the next section are obtained through CSM.

\section{Similarity Measure} \label{section5}
This section investigates the impact of factors such as deployment altitude and area on relative error through numerical results, thereby exploring the necessity of spherical modeling and the feasibility of planar approximation. We consider the channel follows the aerial-to-ground channel model given in \cite{alzenad2019coverage} when $h_s = R_s - R_{\oplus} \leq 1000$ km, while the channel follows the space-to-ground channel model given in \cite{talgat2020stochastic} when $h_s \geq 10,000$~km. Since the authors in \cite{talgat2020stochastic} did not consider interference, we assume that links between interfering NTPs and the typical user follow the same space-to-ground fading model with a uni-directional antenna gain (only the transmitting antenna gain is considered). Considering that the primary focus of this article is NTP, the issue we investigate is the choice between planar and spherical modeling for NTN in uplink communications. {\color{black}Unless otherwise specified, we assume $N_p = 20$ NTPs follow a  homogeneous BPP in this section, and the positions of NTPs are determined by Algorithm~\ref{alg1}.}

\subsection{Deployment Altitude}
\begin{figure}[ht]
\centering
\includegraphics[width=0.8\linewidth]{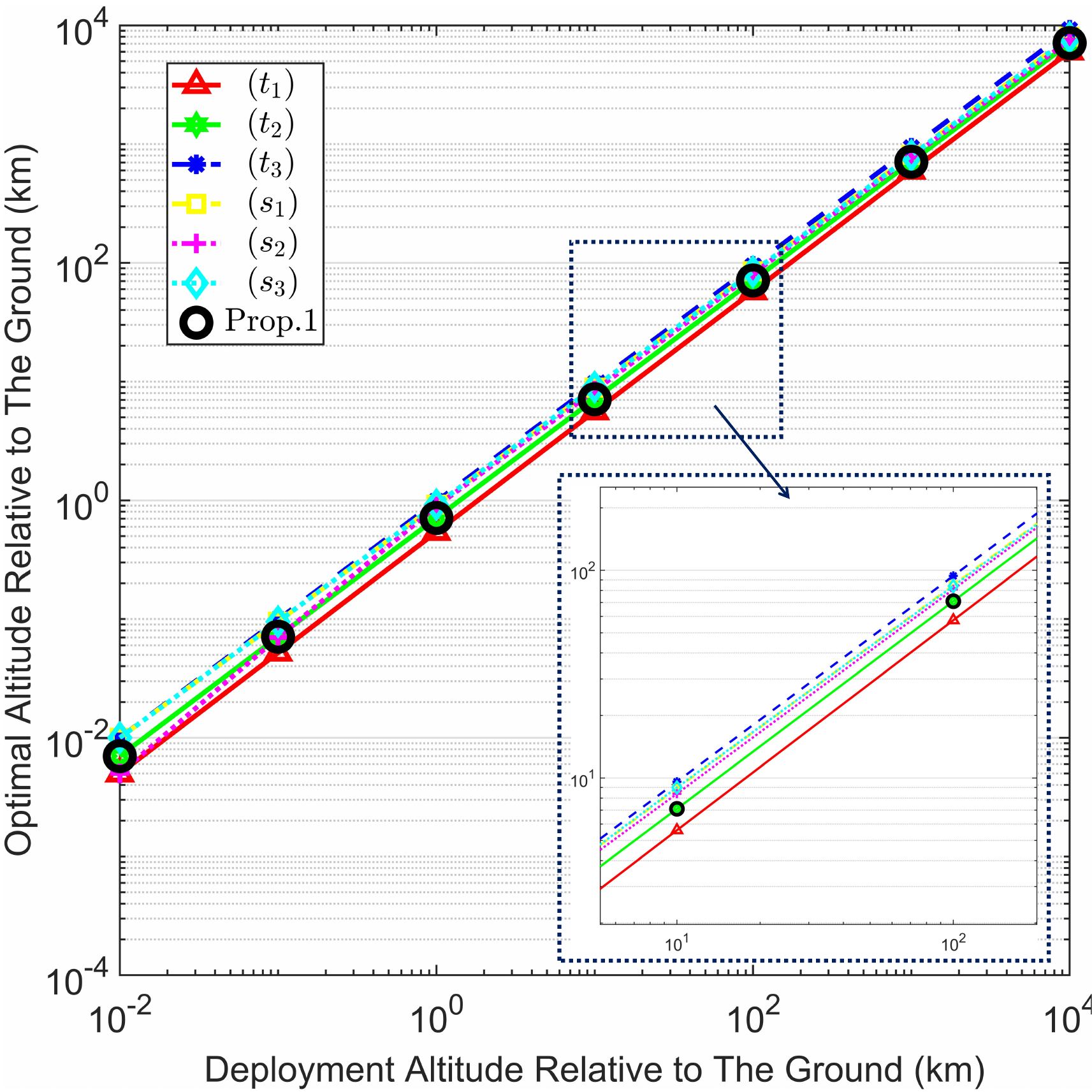}
\caption{Results on optimal deployment altitudes.}
\label{figure2}
\end{figure}

In Fig.~\ref{figure2}, we verify the accuracy of the altitude derived in Proposition~\ref{prop1}, as well as the feasibility of replacing the outer loop of Algorithm~\ref{alg2} with this altitude. The $x$-axis in Fig.~\ref{figure2} is deployment altitude relative to the ground, which is calculated by $h_s = R_s-R_{\oplus}$. The $y$-axis is $h_{\mathrm{opt}}-R_{\oplus}$, where $h_{\mathrm{opt}}$ of the metric is obtain by Algorithm~\ref{alg2}. NTPs are deployed in the user's line-of-sight (LoS) range, which is the spherical cap with $\theta_{\max} = \arccos \left( \frac{R_{\oplus}}{R_s} \right)$.

\par
As shown in Fig.~\ref{figure2}, the optimal altitude given by Proposition~\ref{prop1} overlaps well with that of $(t_2)$, which proves the expression in Proposition~\ref{prop1} is accurate. Additionally, the optimal altitudes corresponding to system-level metrics are generally higher than those of topology-related metrics. Overall, the impact of metric selection on the optimal altitude is relatively limited. The average optimal altitude of the six parameters is $1.085$ times the $h_{\mathrm{opt}}$ given in Proposition~\ref{prop1}. 
Although the estimated altitude provided in Proposition~\ref{prop1} is slightly underestimated, this deviation is relatively acceptable. 

\begin{figure}[ht]
\centering
\includegraphics[width=0.8\linewidth]{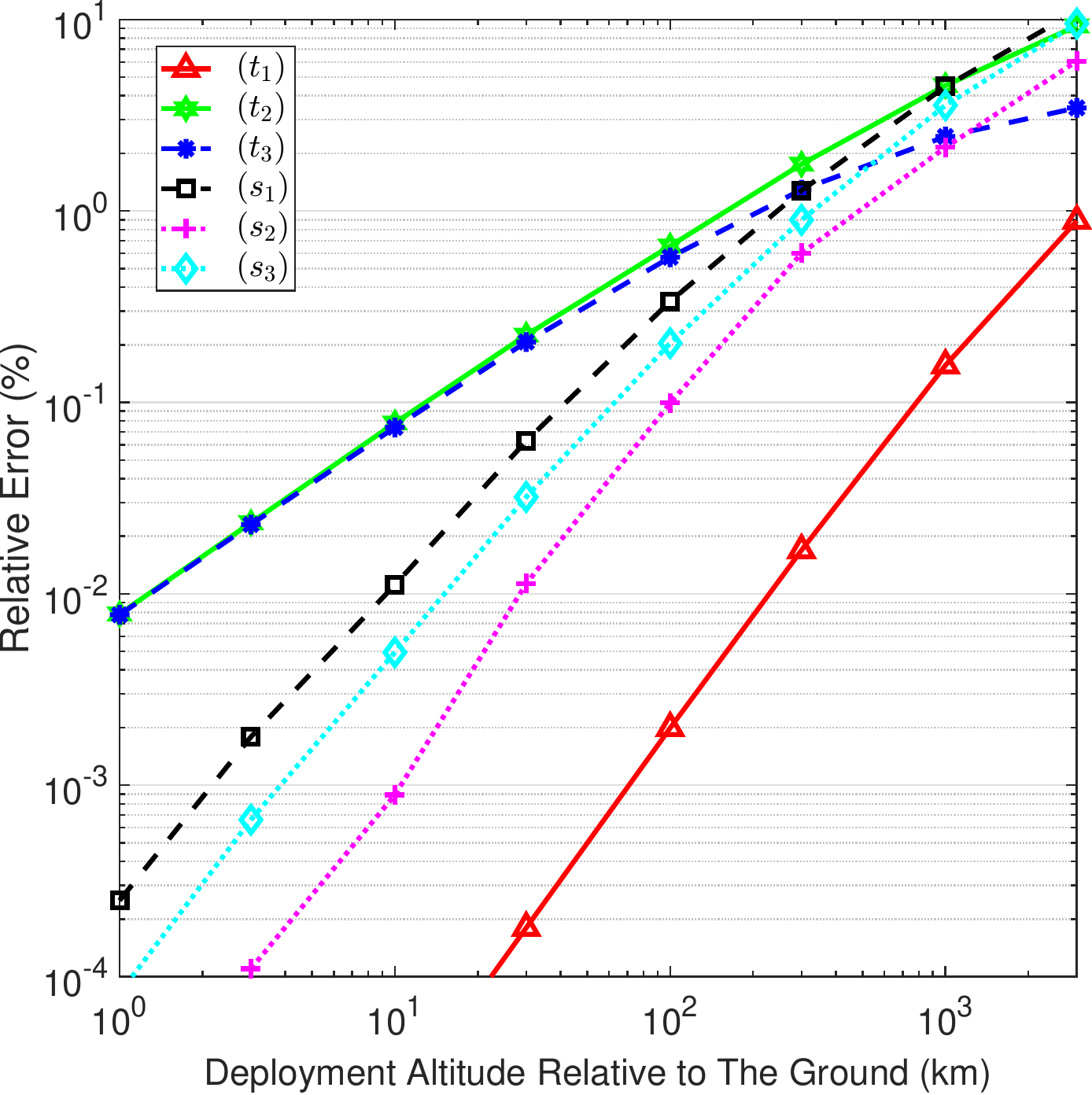}
\caption{Relative error with different altitudes and metrics.}
\label{figure3}
\end{figure}

\par
Next, Fig.~\ref{figure3} explores the impact of the NTN deployment altitude on the relative error. Note that the $y$-axis is in units of $\%$, for example, $10^0\% = 0.01$. And $\theta_{\max} = \arccos \left( \frac{R_{\oplus}}{R_s} \right)$ is satisfied. Fig.~\ref{figure3} presents a rather unsurprising conclusion: as the deployment altitude increases, the relative error rises. In addition, the choice of metric significantly impacts the relative error. Even when studying the same NTN, planar modeling may yield an acceptable relative error for one metric, whereas spherical modeling may be necessary for another.

\subsection{Deployment Region}
Next, the results of relative errors with different deployment regions and altitudes are presented by heat maps. First, we introduce two conditions for the deployment region. 
\begin{itemize}
    \item Beam angle: The ground user is equipped with an antenna, with the beam directed towards the user's zenith. The central angle of the antenna's main lobe beamwidth is fixed as $\psi$, and NTPs falling within the main lobe can communicate with the typical user. 
    \item Area: The user can communicate with NTPs located within a spherical cap of fixed area $\mathcal{A}$ at the user's zenith.
\end{itemize}

\par
Note that the above conditions refer to spherical modeling. For a given spherical model, Algorithm~\ref{alg1} and Algorithm~\ref{alg2} uniquely determine the corresponding planar model. To facilitate the simulation implementation, the following lemma derives the relationship between $\psi$, $\mathcal{A}$, and $\theta_{\max}$.

\begin{lemma}\label{lemma2}
Given the central angle of the antenna's main lobe beamwidth as $\psi$, the maximum central angle is
\begin{equation}
\begin{split}
    \theta_{\max} = \arccos \Bigg( & \frac{R_s^2 + R_{\oplus}^2}{2 R_s R_{\oplus}} - \frac{1}{2 R_s R_{\oplus}} \Bigg( R_{\oplus} \cos\left(\pi - \frac{\psi}{2} \right) \\
    & + \sqrt{R_{\oplus}^2 \cos^2\left(\pi - \frac{\psi}{2} \right) + R_s^2 - R_{\oplus}^2} \bigg)^2 \Bigg).
\end{split}
\end{equation}
Given the area is $\mathcal{A}$, the maximum central angle is
\begin{equation}
    \theta_{\max} = \arccos \left( 1 - \frac{\mathcal{A}}{2\pi R_s^2} \right).
\end{equation}
\end{lemma}
\begin{IEEEproof}
See Appendix~\ref{app:lemma2}.
\end{IEEEproof}

\begin{figure}[ht]
\centering
\includegraphics[width=0.9\linewidth]{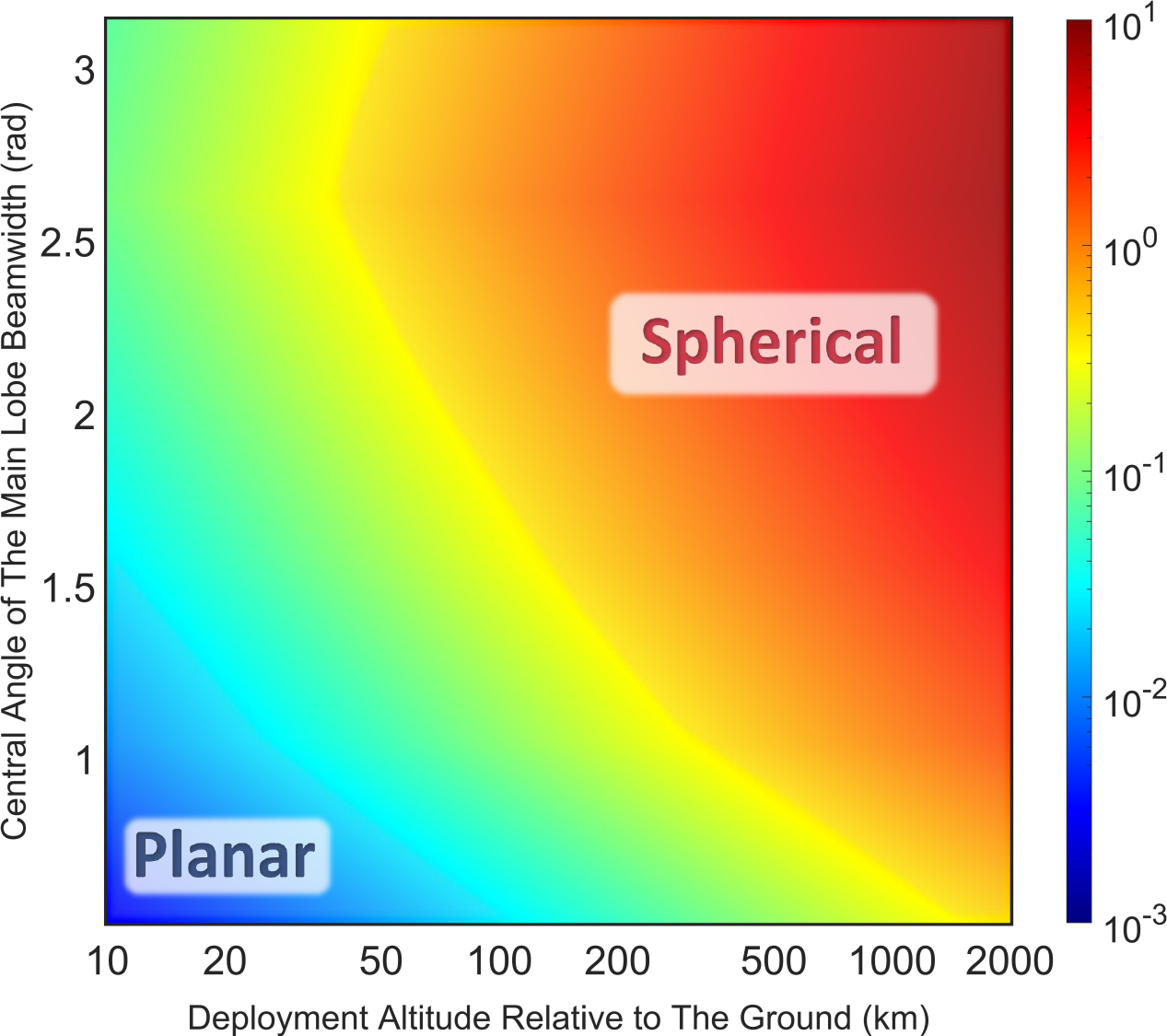}
\caption{Relative error with different altitudes and deployment regions (beam angle).}
\label{figure4}
\end{figure}

\begin{figure}[ht]
\centering
\includegraphics[width=0.9\linewidth]{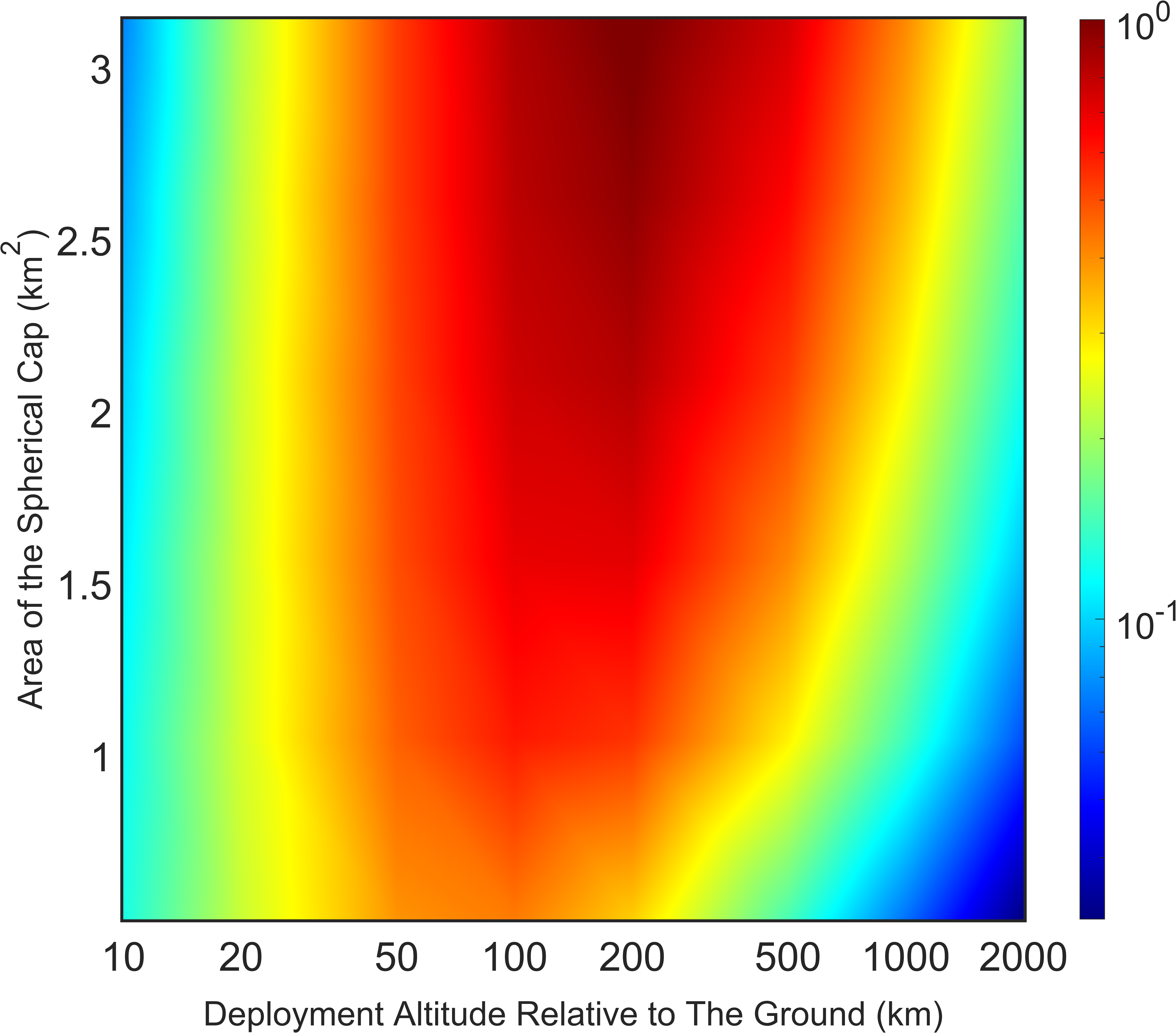}
\caption{Relative error with different altitudes and deployment regions (area).}
\label{figure5}
\end{figure}

{\color{black}In Fig.~\ref{figure4} and Fig.~\ref{figure5}, the relative error with metric $(t_2)$ is shown. Note that the relative errors in these two figures are also expressed as a percentage. For example, in Fig.~\ref{figure4}, the red color represents 10\%, not 10.} Fig.~\ref{figure4} demonstrates the complementary effect between altitude $h_s$ and the beam main lobe central angle $\psi$. Only users equipped with highly directional beams can apply planar modeling for higher-altitude NTN without introducing significant modeling errors. Additionally, when the acceptable thresholds for relative error are determined, the heat map can visually show the bias for modeling approaches. Taking Fig.~\ref{figure4} as an example, if the parameter pair $(h_s,\psi)$ falls within the red region, spherical modeling is necessary. If $(h_s,\psi)$ falls within the blue region, planar modeling is recommended to simplify the analysis. For the intermediate region, there is no clear preference between the two modeling approaches.

\par
As the $y$-axis shifts from $\psi$ to the area of the spherical cap $\mathcal{A}$, Fig.~\ref{figure5} exhibits a completely different trend compared to Fig.~\ref{figure4}. Firstly, when the altitude $h_s$ is fixed, the effect of $\mathcal{A}$ on relative error is uncertain. In most cases, as $\mathcal{A}$ increases, the relative error also increases. Secondly, when $\mathcal{A}$ is fixed, the relative error is highest when $h_s$ is between $100$~km and $200$~km. These findings highlight the complexity of the choice of planar and spherical modeling, as well as the importance of quantitatively analyzing the relative error.

\subsection{Case Study}
This subsection analyzes the relative error under different $\psi$ using HAPs ($h_s = 20$~km) and LEO satellites ($h_s = 550$~km) as case studies. For both cases, as $\psi$ decreases linearly, the relative error decreases exponentially, highlighting the strong influence of directional reception on modeling choices. 

\begin{figure}[ht]
\centering
\includegraphics[width=0.8\linewidth]{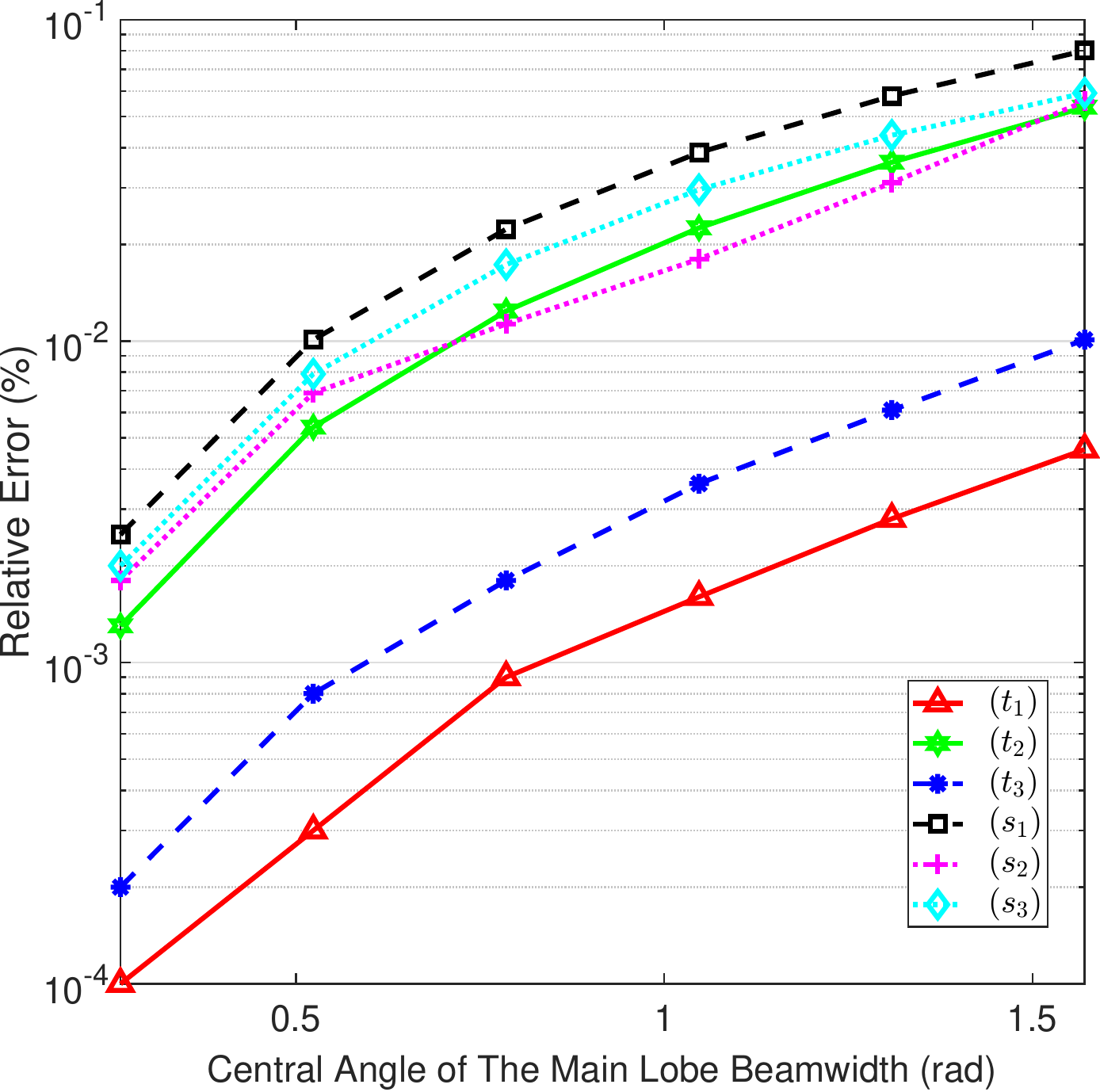}
\caption{Relative error for HAPs ($h_s = 20$~km).}
\label{figure6}
\end{figure}

\par
When users are equipped with directional receiving antennas, the relative error results for typical system-level parameters are relatively close. Therefore, we use them as references to provide recommendations for modeling choices. Compared to HAPs, the relative error of LEO satellites is greater by more than one order of magnitude.

\begin{figure}[ht]
\centering
\includegraphics[width=0.8\linewidth]{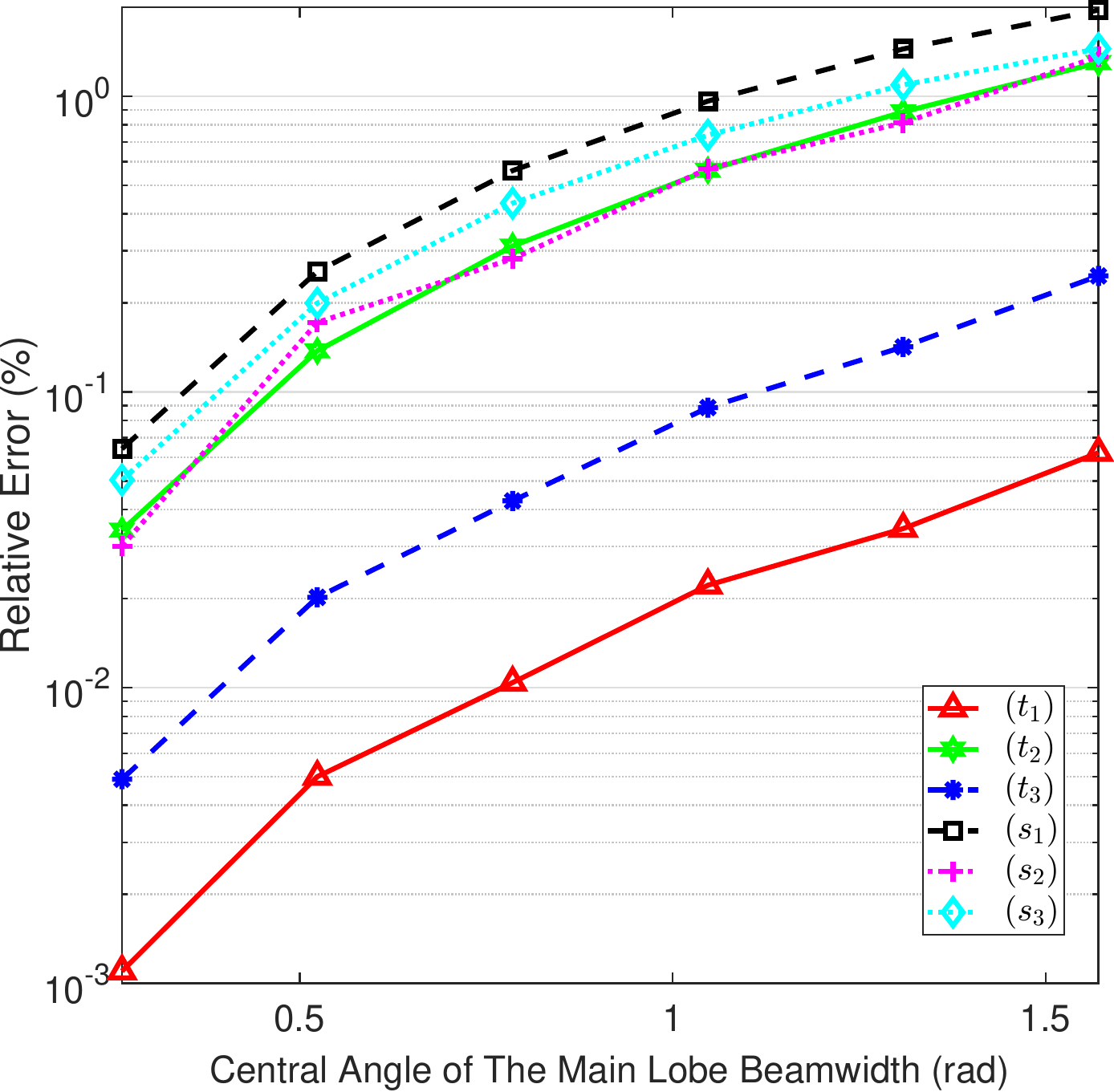}
\caption{Relative error for LEO satellites ($h_s = 550$~km).}
\label{figure7}
\end{figure}

We assume that $0.1\%$ is an acceptable threshold for relative error. For HAP, even with a wider beam main lobe, such as $\psi = \frac{\pi}{2}$, planar modeling remains the preferred choice. {\color{black}In contrast, for LEO satellites, if equipped with a relatively wider beam, such as $\psi > \frac{\pi}{12}$, spherical modeling is necessary.}

\section{Conclusion and Future Work}
This article investigated the similarity measurement between planar and spherical models in the field of SG. Several existing studies have mapped spherical models to planar models for the sake of simplification in analysis. This article proposed a mapping scheme that ensures the communication distances of the mapped planar point process do not exhibit an overall underestimation or overestimation. Furthermore, the valuable mathematical property of homogeneity can be preserved. 

\par
Regarding the choice between planar and spherical modeling, we provide a quantifiable measurement scheme based on relative error. The modeling selection is influenced by numerous factors, and the relationship between factors such as deployment area and the modeling selection is not intuitive. However, in most cases, the lower the deployment altitude and the smaller the deployment area, the more suitable it is to apply planar modeling. {\color{black}For HAPs, planar modeling can often be prioritized as the preferred choice. 

\par
So far, the discussion scope has focused on observing the NTN from the user's perspective. The authors in \cite{wang2023reliability} applied the SG framework to the performance analysis of multi-layer NTNs. In this case, multiple spherical caps can be individually mapped to multiple corresponding planes, and the overall relative error can be represented as a weighted sum of the single-layer relative errors (the weights may depend on factors such as density). Furthermore, the authors in \cite{wang2024ultra} analyzed the performance of inter-satellite links in satellite routing using the SG framework. In this work, relay satellites select the next-hop from the spherical cap region around the ideal relay position, which corresponds to observing the spherical PP from the perspective of a point inside the PP. Here, the nearby satellites can be projected onto a plane by treating the ideal relay position as the vertex of a spherical cap, and the method for calculating the relative error remains unchanged. However, from the perspective of the previous-hop satellite, the mapping corresponds to projecting a non-standard (irregular) spherical cap onto an elliptical plane. The relative error results and mapping strategies under the above two scenarios can be left for future work.
}

\appendices
\section{Proof of Theorem~\ref{theorem1}} \label{app:theorem1}
We begin by proving the homogeneity of $\mathcal{X}$. As the radius of the spherical cap is fixed as $R_s$, we will prove that the probability of a point $x^{(n)}$ locates within region $\varphi_0 < \varphi_s^{(n)} < \varphi_0 + \Delta \varphi$, $\theta_0 < \theta_s^{(n)} < \theta_0 + \Delta \theta$ is directly proportional to the area of the region, where $1 \leq n \leq N_p$. $\Delta \varphi$ and $\Delta \theta$ representing differential elements of the angle are positive values infinitesimally close to $0$. For the azimuth angle,
\begin{equation}\label{AppA-1}
\begin{split}
  &\mathbbm{P} \left[ \varphi_0 < \varphi_s^{(n)} < \varphi_0 + \Delta \varphi \right] \\ & = \mathbbm{P} \left[ \frac{\varphi_0}{2\pi} < v^{(n)} < \frac{\varphi_0 + \Delta \varphi}{2\pi} \right] = \frac{\Delta \varphi}{2\pi},
\end{split}
\end{equation}
where $v^{(n)}$ is uniformly distributed between $0$ and $1$. For the polar angle,
\begin{equation}\label{AppA-2}
\begin{split}
    & \mathbbm{P} \left[ \theta_0 < \theta_s^{(n)} < \theta_0 + \Delta \theta \right] \\ & = \mathbbm{P} \left[ \frac{ 1- \cos\theta_0}{1- \cos\theta_{\max}} < u^{(n)} < \frac{ 1- \cos(\theta_0+\Delta\theta)}{1- \cos\theta_{\max}} \right] \\ 
    & = \frac{ \cos\theta_0 - \cos(\theta_0 + \Delta\theta)} {1- \cos\theta_{\max}} \overset{(a)}{=} \frac{\sin\theta_0 \Delta\theta}{1- \cos\theta_{\max}},
\end{split}
\end{equation}
where $u^{(n)}$ is uniformly distributed between $0$ and $1$, and step $(a)$ follows the definition of the derivative of the cosine function. The probability of $x^{(n)}$ locates within region $\varphi_0 < \varphi_s^{(n)} < \varphi_0 + \Delta \varphi$, $\theta_0 < \theta_s^{(n)} < \theta_0 + \Delta \theta$ is
\begin{equation}\label{AppA-3}
\begin{split}
    & \mathbbm{P} \left[ \varphi_0 < \varphi_s^{(n)} < \varphi_0 + \Delta \varphi, \theta_0 < \theta_s^{(n)} < \theta_0 + \Delta \theta \right] \\
    & \overset{(b)}{=} \mathbbm{P} \left[ \varphi_0 < \varphi_s^{(n)} < \varphi_0 + \Delta \varphi\right] \\ & \times \mathbbm{P} \left[ \theta_0 < \theta_s^{(n)} < \theta_0 + \Delta \theta \right] = \frac{\sin\theta_0 \Delta\theta \Delta\varphi}{2 \pi (1- \cos\theta_{\max})},
\end{split}
\end{equation}
where step (b) follows the independence of the distributions of $u^{(n)}$ and $v^{(n)}$. Since the area of region $\varphi_0 < \varphi_s^{(n)} < \varphi_0 + \Delta \varphi$, $\theta_0 < \theta_s^{(n)} < \theta_0 + \Delta \theta$ is $R_s^2 \sin \theta_0 \Delta \theta \Delta \varphi$ and $R_s$ is a constant, $\mathcal{X}$ is homogeneous.

\par
The proof of homogeneity of $\mathcal{Y}$ follows similar procedures. The probability that a point $y^{(n)}$ locates within region $\rho_0 < \rho_p^{(n)} < \rho_0 + \Delta \rho$, $\varphi_0 < \varphi_p^{(n)} < \varphi_0 + \Delta \varphi$ is,
\begin{equation}\label{AppA-4}
\begin{split}
    & \mathbbm{P} \left[ \rho_0 < \rho_p^{(n)} < \rho_0 + \Delta \rho, \varphi_0 < \varphi_p^{(n)} < \varphi_0 + \Delta \varphi \right] \\
    & = \mathbbm{P} \left[ \rho_0 < \rho_p^{(n)} < \rho_0 + \Delta \rho \right] \\
    & \times \mathbbm{P} \left[\varphi_0 < \varphi_p^{(n)} < \varphi_0 + \Delta \varphi \right] 
    \\ & = \mathbbm{P} \left[ 
    \frac{\rho_0^2}{\rho_{\max}^2} <  u^{(n)} < \frac{ \left(\rho_0+\Delta \theta \right)^2}{\rho_{\max}^2} \right] \times \frac{\Delta\varphi}{2\pi} \\
    & = \frac{\rho_0 \Delta\rho \Delta\varphi}{\pi \rho_{\max}^2},
\end{split}
\end{equation}
which is directly proportional to the area of the region $\rho_0 \Delta \rho \Delta \varphi$. Therefore, $\mathcal{Y}$ is also homogeneous.

\section{Proof of Theorem~\ref{theorem2}} \label{app:theorem2}
To begin with, we convert cylindrical coordinates and spherical coordinates into Cartesian coordinates:
\begin{equation}\label{AppB-1}
\begin{split}
    & \left\{
 	\begin{array}{ll}
    \widetilde{x}_s^{(n)} = R_s \sin\theta_s^{(n)} \cos\varphi_s^{(n)}, & \widetilde{x}_p^{(n)} = \rho_p^{(n)}\cos\varphi^{(n)}, \\
    \widetilde{y}_s^{(n)} = R_s \sin\theta_s^{(n)} \sin\varphi_s^{(n)}, & \widetilde{y}_p^{(n)} = \rho_p^{(n)}\sin\varphi^{(n)}, \\
    \widetilde{z}_s^{(n)} = R_s \cos\theta_s^{(n)}, & \widetilde{z}_p = h_p.
	\end{array}
	\right.
\end{split}
\end{equation}
By observing (\ref{AppB-1}), it can be deduced that as $R_s$ tends to infinity, $h_p$ converges to $R_s \cos\theta_s^{(n)}$, and $\rho_p^{(n)}$ converges to $R_s \sin\theta_s^{(n)}$, then points $x^{(n)} \left(R_s,\varphi_s^{(n)},\theta_s^{(n)} \right)$ and $y^{(n)} \left(\rho_p^{(n)},\varphi_p^{(n)},h_p \right)$ coincide. 

\par
Given that $S=2\pi R_s^2 (1-\cos\theta_{\max})$ is a finite value, 
\begin{equation}
\begin{split}
    & R_s \cos\theta_s^{(n)} = R_s \left( 1 - u^{(n)} \left(1-\cos\theta_{\max}\right) \right) \\
    & = R_s \left( 1 - u^{(n)} \frac{S}{2\pi R_s^2} \right) \overset{R_s \rightarrow \infty}{\xlongequal{\quad\quad}} R_s
\end{split}
\end{equation}
satisfies for $\forall \ 0 \leq \theta_s^{(n)} \leq \theta_{\max}$. Therefore, as long as $R_s\cos\theta_{\max} < h_p < R_s$, $h_p$ and $R_s \cos\theta_s^{(n)}$ both converge to $R_s$.

\par
As for the latter convergence, we first focus on $R_s \sin\theta_s^{(n)}$,
\begin{equation}
\begin{split}
    & \left( R_s \sin\theta_s^{(n)} \right)^2 = R_s^2 \left( 1 - u^{(n)} \left(1-\cos\theta_{\max}\right) \right) \\
    & = R_s^2 u^{(n)} \frac{S}{\pi R_s^2} - R_s^2 \left( u^{(n)} \right)^2 \frac{S^2}{\pi^2 R_s^4} \overset{R_s \rightarrow \infty}{\xlongequal{\quad\quad}} \frac{u^{(n)} S}{\pi}.
\end{split}
\end{equation}
When $\rho_{\max} = R_s \cos\theta_{\max}$,
\begin{equation}
\begin{split}
    & \left( \rho_p^{(n)} \right)^2 = u^{(n)} R_s^2 \left( 1 - \left(\cos\theta_{\max}\right)^2 \right) \\ & = u^{(n)} R_s^2 \left( 1 - \cos\theta_{\max} \right) \left( 1 + \cos\theta_{\max} \right) \\
    & = u^{(n)} \frac{S}{2\pi} \left( 1 - \frac{S}{2\pi R_s^2} + 1\right) \overset{R_s \rightarrow \infty}{\xlongequal{\quad\quad}} \frac{u^{(n)} S}{\pi}.
\end{split}
\end{equation}
Since both $R_s \sin\theta_s^{(n)}$ and $\rho_p^{(n)}$ converge to $\sqrt{u^{(n)} S/ \pi}$, $\mathcal{X}$ and $\mathcal{Y}$ are asymptotically similar.

\section{Proof of Lemma~\ref{lemma1}} \label{app:lemma1}
In this appendix, we derive the CDF of the distance distributions for two PPs and determine their range of values. 
The CDF of the distance distribution between an arbitrary NTP $x^{(n)}\left( R_s, \varphi_s^{(n)}, \theta_s^{(n)} \right)$ in the spherical point process $\mathcal{X}$ and the user can be expressed as
\begin{equation}
\begin{split}
    & F_s(d) = \mathbbm{P} \left[ D_s \leq d \right] = \mathbbm{P} \left[ D_s^2 \leq d^2 \right] \\
    & = \mathbbm{P} \left[ R_{\oplus}^2 + R_s^2 - 2R_{\oplus} R_s \cos\theta_s^{(n)} \leq d^2 \right], \\
    & = \mathbbm{P} \left[ \cos\theta_s^{(n)} \geq \frac{R_{\oplus}^2 + R_s^2 - d^2}{2R_{\oplus} R_s}\right] \\
    & \overset{(a)}{=} \mathbbm{P} \left[ 1 - u^{(n)} \left(1 - \cos\theta_{\max} \right) \geq \frac{R_{\oplus}^2 + R_s^2 - d^2}{2R_{\oplus} R_s}\right] \\
    & \overset{(b)}{=} \frac{1}{\cos\theta_{\max}} \left( 1 - \frac{R_{\oplus}^2 + R_s^2 - d^2}{2R_{\oplus} R_s} \right),
\end{split}
\end{equation}
where step $(a)$ is obtained by the expression in step (4) of Algorithm~\ref{alg1}, and step $(b)$ holds since $u^{(n)}$ follows the uniform distribution. 

\par
Then, we determine the range of $d$. When the NTP is directly above the user, i.e., $\theta_s^{(n)}=0$, the distance is shortest, which is $d = R_s - R_{\oplus}$. When condition $\theta_s^{(n)} = \theta_{\max}$ holds, $d$ reaches its maximum value, which is 
\begin{equation}
    d = \sqrt{R_{\oplus}^2 + R_s^2 - 2R_{\oplus} R_s \cos\theta_{\max}}.
\end{equation}

\par
As for the CDF of the distance between an arbitrary NTP $y^{(n)} \left(\rho_p^{(n)},\varphi_p^{(n)},h_p \right)$ and the user, it can be expressed as
\begin{equation}
\begin{split}
    & F_p(d) = \mathbbm{P} \left[ D_p \leq d \right] = \mathbbm{P} \left[ \rho_p^{(n)} \leq \sqrt{d^2 - h_p^2} \right] \\
    & \overset{(c)}{=} \mathbbm{P} \left[ \rho_{\max}^2 u^{(n)} \leq d^2 - h_p^2 \right]  \overset{(d)}{=} \frac{d^2 - h_p^2}{\rho_{\max}^2},
\end{split}
\end{equation}
where step $(c)$ is given in step (4) of Algorithm~\ref{alg1}, and step $(d)$ is obtained since $u^{(n)}$ follows the uniform distribution. When $\rho^{(n)} = 0$ and $\rho^{(n)} \rho_{\max}$, $d$ reaches its minimum value $h_p$ and maximum value $\sqrt{h_p^2 + \rho_{\max}^2}$, respectively.

\section{Proof of Proposition~\ref{prop1}} \label{app:prop1}
Since $h_p$ does not influence that value of $\mathcal{G}(\mathcal{X})$, our task is actually find a $h_{\mathrm{opt}}$ that minimize $\left| \mathcal{G}(\mathcal{X}) - \mathcal{G}(\mathcal{Y}) \right|$. Then, $\mathcal{G}(\mathcal{X})$ and $\mathcal{G}(\mathcal{Y})$ can be expressed as,
\begin{equation}
\begin{split}
    \mathcal{G}(\mathcal{X}) & = \sum_{n=1}^{N_s} R_s^2 + R_{\oplus}^2 \\
    & - 2R_s R_{\oplus} \left( 1 - u^{(n)} (1-\cos\theta_{\max}) \right),
\end{split}
\end{equation}
\begin{equation}
    \mathcal{G}(\mathcal{Y}) = \sum_{n=1}^{N_s} (h_p - R_{\oplus})^2 + u^{(n)} \rho_{\max}^2.
\end{equation}
According to the law of large numbers, when $N_s \rightarrow \infty$, 
\begin{equation}\label{AppC-3}
\begin{split}
& \left| \mathcal{G}(\mathcal{X}) - \mathcal{G}(\mathcal{Y}) \right| = N_s \mathbbm{E}_u \Big[ (h_p - R_{\oplus})^2 + u \rho_{\max}^2 \\ & - R_s^2 - R_{\oplus}^2 + 2 R_s R_{\oplus} \left( 1 - u (1-\cos\theta_{\max}) \right) \Big] \\
& = N_s \int_0^1 \Big( \left(\rho_{\max}^2 - 2 (1-\cos\theta_{\max}) R_s R_{\oplus}
 \right)u \\ & + \left( h_p^2 - 2h_p R_{\oplus} - R_s^2 + 2R_s R_{\oplus} \right) \Big) f_u(u) {\mathrm{d}}u \\
& = N_s \bigg( h_p^2 - 2h_p R_{\oplus} + \bigg( \frac{1}{2}\rho_{\max}^2 - R_s^2 \\ & - (1 -\cos\theta_{\max}) R_s R_{\oplus} + 2R_s R_{\oplus} \bigg) \bigg).
\end{split}
\end{equation}
In this case, we find an optimal $h_{\mathrm{opt}}$ let $\left| \mathcal{G}(\mathcal{X}) - \mathcal{G}(\mathcal{Y}) \right|=0$ given that $h_{\mathrm{opt}} > R_{\oplus}$,
\begin{equation}
\begin{split}
    & h_{\mathrm{opt}} = R_{\oplus} \\
    & + \sqrt{R_{\oplus}^2 - \frac{1}{2}\rho_{\max}^2 - (1+\cos\theta_{\max}) R_s R_{\oplus} + R_s^2}.
\end{split}
\end{equation}

\section{Proof of Lemma~\ref{lemma2}} \label{app:lemma2}
{\color{black}Given that the center of the antenna's main lobe beamwidth is $\psi$, the following equation can be derived using the Cosine theorem,}
\begin{equation}
    \cos \left( \pi - \frac{\psi}{2} \right) = \frac{d_{\max}^2 + R_{\oplus}^2 - R_s^2}{2d_{\max} R_{\oplus}},
\end{equation}
where $d_{\max}$ is the maximum Euclidean distance between the NTP in the main lobe and the typical user. The quadratic equation about $d_{\max}$ can be further derived,
\begin{equation}
    d_{\max}^2 - 2 R_{\oplus} \cos \left( \pi - \frac{\psi}{2} \right) d_{\max} + R_{\oplus}^2 - R_s^2 = 0.
\end{equation}
Since $d_{\max}$ is a positive value, we have
\begin{equation}\label{AppE-3}
\begin{split}
    d_{\max} & = R_{\oplus} \cos \left( \pi - \frac{\psi}{2} \right) \\
    & + \sqrt{R_{\oplus}^2 \cos^2 \left( \pi - \frac{\psi}{2} \right) + R_s^2 - R_{\oplus}^2}.
\end{split}
\end{equation}
Next, the relationship of $\theta_{\max}$ and $d_{\max}$ is
\begin{equation}\label{AppE-4}
    \theta_{\max} = \arccos\left( \frac{ R_{\oplus}^2 + R_s^2 - d_{\max}^2 }{ 2 R_{\oplus} R_s } \right).
\end{equation}
Substitute (\ref{AppE-3}) into (\ref{AppE-4}), $\theta_{\max}$ can be expressed by $\psi$. 

\par
As for the condition of the fixed area, the relationship between  $\mathcal{A}$ and $\theta_{\max}$ is given as
\begin{equation}
    \mathcal{A} = 2 \pi R_s \left( 1 - \cos\theta_{\max} \right),
\end{equation}
and the final result can be obtained through simple transformations.

\bibliographystyle{IEEEtran}
\bibliography{references}

\begin{thebibliography}{10}
\providecommand{\url}[1]{#1}
\csname url@samestyle\endcsname
\providecommand{\newblock}{\relax}
\providecommand{\bibinfo}[2]{#2}
\providecommand{\BIBentrySTDinterwordspacing}{\spaceskip=0pt\relax}
\providecommand{\BIBentryALTinterwordstretchfactor}{4}
\providecommand{\BIBentryALTinterwordspacing}{\spaceskip=\fontdimen2\font plus
\BIBentryALTinterwordstretchfactor\fontdimen3\font minus \fontdimen4\font\relax}
\providecommand{\BIBforeignlanguage}[2]{{%
\expandafter\ifx\csname l@#1\endcsname\relax
\typeout{** WARNING: IEEEtran.bst: No hyphenation pattern has been}%
\typeout{** loaded for the language `#1'. Using the pattern for}%
\typeout{** the default language instead.}%
\else
\language=\csname l@#1\endcsname
\fi
#2}}
\providecommand{\BIBdecl}{\relax}
\BIBdecl

\bibitem{belmekki2022unleashing}
B.~E.~Y. Belmekki and M.-S. Alouini, ``Unleashing the potential of networked tethered flying platforms: Prospects, challenges, and applications,'' \emph{IEEE Open Journal of Vehicular Technology}, vol.~3, pp. 278--320, 2022.

\bibitem{belmekki2024cellular}
B.~E.~Y. Belmekki, A.~J. Aljohani, S.~A. Althubaity, A.~Al~Harthi, K.~Bean, A.~Aijaz, and M.-S. Alouini, ``Cellular network from the sky: {T}oward people-centered smart communities,'' \emph{IEEE Open Journal of the Communications Society}, 2024, early Access.

\bibitem{he2024direct}
Y.~He, Y.~Xiao, S.~Zhang, M.~Jia, and Z.~Li, ``Direct-to-smartphone for {6G NTN}: Technical routes, challenges, and key technologies,'' \emph{IEEE Network}, vol.~38, no.~4, pp. 128--135, 2024.

\bibitem{wang2025modeling}
R.~Wang, M.~A. Kishk, and M.-S. Alouini, ``Modeling and analysis of non-terrestrial networks by spherical stochastic geometry: A survey,'' \emph{IEEE Communications Surveys \& Tutorials}, 2025, early Access.

\bibitem{qiu2023performance}
Z.~Qiu and W.~Wang, ``Performance analysis and simulation of large-scale {LEO} constellation under a stochastic geometric perspective,'' in \emph{ICC 2023-IEEE International Conference on Communications}.\hskip 1em plus 0.5em minus 0.4em\relax IEEE, 2023, pp. 3425--3431.

\bibitem{wang2022ultra}
R.~Wang, M.~A. Kishk, and M.-S. Alouini, ``Ultra-dense {LEO} satellite-based communication systems: {A} novel modeling technique,'' \emph{IEEE Communications Magazine}, vol.~60, no.~4, pp. 25--31, 2022.

\bibitem{gao2019spectrum}
Z.~Gao, Z.~Wei, Z.~Wang, J.~Zhu, G.~Deng, and Z.~Feng, ``Spectrum sharing for high altitude platform networks,'' in \emph{International Conference on Communications in China (ICCC)}.\hskip 1em plus 0.5em minus 0.4em\relax IEEE, 2019, pp. 411--415.

\bibitem{alzenad2019coverage}
M.~Alzenad and H.~Yanikomeroglu, ``Coverage and rate analysis for vertical heterogeneous networks ({VHetNets}),'' \emph{IEEE Transactions on Wireless Communications}, vol.~18, no.~12, pp. 5643--5657, 2019.

\bibitem{zhang2021stochastic}
X.~Zhang, B.~Zhang, K.~An, G.~Zheng, S.~Chatzinotas, and D.~Guo, ``Stochastic geometry-based analysis of cache-enabled hybrid satellite-aerial-terrestrial networks with non-orthogonal multiple access,'' \emph{IEEE Transactions on Wireless Communications}, vol.~21, no.~2, pp. 1272--1287, 2021.

\bibitem{tian2023satellite}
Y.~Tian, G.~Pan, H.~ElSawy, and M.-S. Alouini, ``Satellite-aerial communications with multi-aircraft interference,'' \emph{IEEE Transactions on Wireless Communications}, vol.~22, no.~10, pp. 7008--7024, 2023.

\bibitem{okati2022nonhomogeneous}
N.~Okati and T.~Riihonen, ``Nonhomogeneous stochastic geometry analysis of massive {LEO} communication constellations,'' \emph{IEEE Transactions on Communications}, vol.~70, no.~3, pp. 1848--1860, 2022.

\bibitem{matracia2023uav}
M.~Matracia, M.~A. Kishk, and M.-S. Alouini, ``{UAV}-aided post-disaster cellular networks: A novel stochastic geometry approach,'' \emph{IEEE Transactions on Vehicular Technology}, vol.~72, no.~7, pp. 9406--9418, 2023.

\bibitem{232306}
Z.~Lou, B.~E. Youcef~Belmekki, and M.-S. Alouini, ``{HAPS} in the non-terrestrial network nexus: Prospective architectures and performance insights,'' \emph{IEEE Wireless Communications}, vol.~30, no.~6, pp. 52--58, 2023.

\bibitem{232304}
Z.~Wei, L.~Wang, Z.~Gao, H.~Wu, N.~Zhang, K.~Han, and Z.~Feng, ``Spectrum sharing between high altitude platform network and terrestrial network: Modeling and performance analysis,'' \emph{IEEE Transactions on Communications}, vol.~71, no.~6, pp. 3736--3751, 2023.

\bibitem{242302}
M.~A. Bliss, F.~J. Block, T.~C. Royster, C.~G. Brinton, and D.~J. Love, ``Orchestrating heterogeneous {NTNs}: From stochastic geometry to resource allocation,'' \emph{IEEE Transactions on Aerospace and Electronic Systems}, 2024, early Access.

\bibitem{sun2024performance}
Y.~Sun and R.~Li, ``Performance analysis of satellite-terrestrial integrated radio access networks based on stochastic geometry,'' 2024, available online: https://arxiv.org/abs/2404.09506.

\bibitem{hu2024performance}
X.~Hu, B.~Lin, X.~Lu, P.~Wang, N.~Cheng, Z.~Yin, and W.~Zhuang, ``Performance analysis of end-to-end {LEO} satellite-aided shore-to-ship communications: A stochastic geometry approach,'' \emph{IEEE Transactions on Wireless Communications}, 2024, early Access.

\bibitem{yastrebova2020theoretical}
A.~Yastrebova, I.~Angervuori, N.~Okati, M.~Vehkaper{\"a}, M.~H{\"o}yhty{\"a}, R.~Wichman, and T.~Riihonen, ``Theoretical and simulation-based analysis of terrestrial interference to {LEO} satellite uplinks,'' in \emph{Global Communications Conference (GLOBECOM)}.\hskip 1em plus 0.5em minus 0.4em\relax IEEE, 2020, pp. 1--6.

\bibitem{al2021tractable}
A.~Al-Hourani, ``A tractable approach for predicting pass duration in dense satellite networks,'' \emph{IEEE Communications Letters}, vol.~25, no.~8, pp. 2698--2702, 2021.

\bibitem{okati2020downlink}
N.~Okati, T.~Riihonen, D.~Korpi, I.~Angervuori, and R.~Wichman, ``Downlink coverage and rate analysis of low {E}arth orbit satellite constellations using stochastic geometry,'' \emph{IEEE Transactions on Communications}, vol.~68, no.~8, pp. 5120--5134, 2020.

\bibitem{choi2024modeling}
C.-S. Choi, ``Modeling and analysis of downlink communications in a heterogeneous {LEO} satellite network,'' \emph{IEEE Transactions on Wireless Communications}, 2024, early Access.

\bibitem{jung2023modeling}
D.-H. Jung, H.~Nam, J.~Choi, and D.~J. Love, ``Modeling and analysis of {GEO} satellite networks,'' 2023, available online: https://arxiv.org/abs/2312.15924.

\bibitem{wang2022evaluating}
R.~Wang, M.~A. Kishk, and M.-S. Alouini, ``Evaluating the accuracy of stochastic geometry based models for {LEO} satellite networks analysis,'' \emph{IEEE Communications Letters}, vol.~26, no.~10, pp. 2440--2444, 2022.

\bibitem{al2021session}
A.~Al-Hourani, ``Session duration between handovers in dense {LEO} satellite networks,'' \emph{IEEE Wireless Communications Letters}, vol.~10, no.~12, pp. 2810--2814, 2021.

\bibitem{angervuori2025meta}
I.~Angervuori, M.~Haenggi, and R.~Wichman, ``Meta distribution of the {SIR} in a narrow-beam {LEO} uplink,'' \emph{IEEE Transactions on Communications}, 2025, early Access.

\bibitem{feller1991introduction}
W.~Feller, \emph{An introduction to probability theory and its applications, Volume 2}.\hskip 1em plus 0.5em minus 0.4em\relax John Wiley \& Sons, 1991, vol.~81.

\bibitem{wang2024ultra}
R.~Wang, M.~A. Kishk, and M.-S. Alouini, ``Ultra reliable low latency routing in {LEO} satellite constellations: A stochastic geometry approach,'' \emph{IEEE Journal on Selected Areas in Communications}, vol.~42, no.~5, pp. 1231--1245, 2024.

\bibitem{rubner1998metric}
Y.~Rubner, C.~Tomasi, and L.~J. Guibas, ``A metric for distributions with applications to image databases,'' in \emph{Sixth international conference on computer vision}.\hskip 1em plus 0.5em minus 0.4em\relax IEEE, 1998, pp. 59--66.

\bibitem{talgat2020nearest}
A.~Talgat, M.~A. Kishk, and M.-S. Alouini, ``Nearest neighbor and contact distance distribution for binomial point process on spherical surfaces,'' \emph{IEEE Communications Letters}, vol.~24, no.~12, pp. 2659--2663, 2020.

\bibitem{talgat2020stochastic}
------, ``Stochastic geometry-based analysis of {LEO} satellite communication systems,'' \emph{IEEE Communications Letters}, vol.~25, no.~8, pp. 2458--2462, 2021.

\bibitem{wang2023reliability}
R.~Wang, M.~A. Kishk, and M.-S. Alouini, ``Reliability analysis of multi-hop routing in multi-tier {LEO} satellite networks,'' \emph{IEEE Transactions on Wireless Communications}, vol.~23, no.~3, pp. 1959--1973, 2023.

\end{thebibliography}

\end{document}